\documentclass[eqsecnum,amsmath,preprintnumbers,nofootinbib,aps,11pt]{revtex4} 
\usepackage{epsfig}
\usepackage{amssymb}
\usepackage{subfigure}
\usepackage{hyperref}
\usepackage{graphicx}
\usepackage{enumerate}
\usepackage[countmax]{subfloat}
\usepackage{MnSymbol}

\DeclareMathAlphabet{\mathpzc}{OT1}{pzc}{m}{it}

\def\beq{\begin{equation}}
\def\eeq{\end{equation}}

\def\bea{\arraycolsep .1em \begin{eqnarray}}
\def\eea{\end{eqnarray}}
\def\Tr{{\rm Tr}}

\def\eq#1{(\ref{#1})}

\def\s0#1#2{\mbox{\small{$ \frac{#1}{#2} $}}}
\def\0#1#2{\frac{#1}{#2}}

\def\grgl{\:\hbox to -0.2pt{\lower2.5pt\hbox{$\sim$}\hss}{\raise3pt\hbox{$>$}}\:}
\def\klgl{\:\hbox to -0.2pt{\lower2.5pt\hbox{$\sim$}\hss}{\raise3pt\hbox{$<$}}\:}

\begin{document}

 \title{Asymptotic safety and the cosmological constant}

\author{Kevin Falls}

\affiliation{Instit\"{u}t f\"{u}r Theoretische Physik, Universit\"{a}t Heidelberg, Philosophenweg 16, 69120 Heidelberg, Germany}

\begin{abstract}
We study the non-perturbative renormalisation of quantum gravity in four dimensions. Taking care to disentangle physical degrees of freedom, we observe the topological nature of conformal fluctuations arising from the functional measure. The resulting beta functions possess an asymptotically safe fixed point with a global phase structure leading to classical general relativity for positive, negative or vanishing cosmological constant. If only the conformal fluctuations are quantised we find an asymptotically safe fixed point predicting a vanishing cosmological constant on all scales. At this fixed point we reproduce the critical exponent, $\nu = 1/3$, found in numerical lattice studies by Hamber. This suggests the fixed point may be physical while solving the cosmological constant problem.

\end{abstract}

\date{\today}
\maketitle
\newpage
\tableofcontents

\newpage
\section{Introduction}
Quantum gravity aims to combine the principles of quantum mechanics with the theory of gravity proposed by Einstein nearly a century ago. This classical theory, general relativity, is based on the equivalence principle for all observers.
The theory is described by the Einstein field equations for the metric tensor $g_{\mu\nu}$, which are generally covariant under arbitrary coordinate transformations. In the absence of matter these equations imply that the scalar curvature is given by $R= 4 \Lambda$, where $\Lambda$ is the cosmological constant, and that the theory describes a spin-two fluctuation corresponding to the graviton. 
However, quantum gravity runs into severe difficulties when standard perturbative methods are applied. In particular the theory is perturbatively non-renormalisable already at one loop, in the presence of matter \cite{'tHooft:1974bx}, and at two loops for pure gravity \cite{Goroff:1985th}. This leaves the possibility that gravity can be quantised non-perturbativley. Alternatively one must go beyond general relativity alone by adopting new degrees of freedom and/or symmetry principles.

Another conundrum of quantum gravity relates to the cosmological constant $\Lambda$. The standard folklore is that the cosmological constant is predicted to be of order the Planck scale $M_{\rm Pl}^2= G_N^{-1}$ where $G_N$ is Newton's constant  (here and throughout we use  units $\hbar = 1 = c$). 
Such a prediction comes from naturalness arguments assuming that its value is set by Planck scale physics. On the other hand this reasoning is in contradiction with observation \cite{RevModPhys.61.1}. Indeed, assuming that $\Lambda$ is responsible for the late time acceleration of the universe, the measured value of $G_N \cdot \Lambda$ is some $122$ orders of magnitude less than this prediction. Thus the standard $\Lambda$CDM-model of cosmology is called into question since it suffers from an apparent fine tuning problem for $\Lambda$. 

One possibility is that $\Lambda$ is exactly zero and that the acceleration of the universe comes from another source of dark energy or modified gravity.  
This would imply that flat Minkowski spacetime is the true vacuum of quantum gravity. That this is the case has been conjectured in \cite{Mazur:1989by} where a careful handling of conformal fluctuations $g_{\mu\nu} \to e^{2 \sigma} g_{\mu\nu}$ has been stressed. Furthermore in \cite{Hamber:2013rb} it has been argued that $\Lambda$ should not receive quantum corrections at all since it can always be set to unity by a conformal field redefinition of the metric tensor.
  
Conformal modes also cause a problem for the quantisation of gravity since they make the na\"{i}vely Wick rotated Euclidean action unbounded from below.  
On the other hand the conformal fluctuations are non-dynamical in general relativity. Therefore such apparently pathological fluctuations of $\sigma$ are only influential off-shell or in the presence of matter. In \cite{Mazur:1989by} the correct treatment of the conformal mode has been derived at the semi-classical level. There it was observed that the proper Wick rotation of $\sigma$ ensures that the action is bounded from below, while the dynamics of $\sigma$ are cancelled by a Jacobian arising in the functional measure.

Ultimately to understand the stability of gravity with or without a cosmological constant we must appeal to the full quantum theory. After quantisation the classical action $S[\varphi]$ of a theory is replaced by the effective action $\Gamma[\phi]$, which results from a Legendre transform of the functional integral. This implies that the effective action is a convex functional of the mean field $\phi = \langle \varphi \rangle$ such that its second functional derivative is positive definite
\beq \label{ConvexGamma}
\Gamma^{(2)}[\phi] >0\,.
\eeq
This condition reflects the stability of the theory and allows for the determination of the vacuum state.
 If we wish to quantise gravity as a fundamental theory this necessitates that we compute $\Gamma[\phi]$ via non-perturbative methods. Making sure \eq{ConvexGamma} continues to be satisfied when approximations are applied is therefore crucial for their consistency. 
At a technical level these considerations relate directly to the regulated functional measure of the path integral and therefore to how the gauge fixing and renormalisation schemes are implemented.

In this paper we shall investigate the non-perturbative quantisation of gravity at an ultra-violet (UV) fixed  point of the renormalisation group (RG) \cite{Weinberg:1980gg}, corresponding to a second order phase transition for quantum gravity. A theory defined at such a fixed point is said to be asymptotically safe provided the phase transition has finitely many relevant directions. In light of the above considerations we shall pay particular attention to the treatment of the cosmological constant, conformal fluctuations and ultimately the convexity condition \eq{ConvexGamma}. While we study a simple phase diagram, parameterised by only the Newtons coupling and the cosmological constant, we shall close the approximation scheme by a non-perturbative expansion ensuring that the effective action remains convex. 
In this way we aim to minimise unphysical contributions while capturing the physics of quantum general relativity namely the spin-two fluctuations of the graviton and the topological conformal modes.

 Aside from asymptotic safety it has been suggested \cite{Hooft:2010ac} that gravity could be quantised by first integrating out the conformal fluctuations and 
then obtaining a conformally invariant effective theory for the remaining degrees of freedom. Then, due to its conformal nature, one would expect the resulting theory 
to remain finite after further quantisation. These ideas came from observing that `complementary' descriptions of evaporating black holes are related by conformal transformations \cite{Hooft:2009ms}.  The problem with this approach is that the conformal modes remain power counting non-renormalisable \cite{Hooft:2010ac}.  
Therefore, the existence of an asymptotically safe UV fixed point for the conformal fluctuations would be desirable. Indeed an asymptotically safe fixed point implies that the theory becomes scale invariant at short distances and that small black hole horizons admit conformal scaling laws \cite{Falls:2012nd}.
In addition to full quantum gravity, we shall therefore investigate the conformally reduced theory where only the conformal modes are quantised.

The rest of this paper is as follows. First we review the functional renormalisation group for gravity and the asymptotic safety scenario in section~\ref{secRG}. In section \ref{dof} we consider the physical and propagating degrees of freedom in quantum general relativity. We adopt a gauge fixing procedure which makes the nature of these degrees of freedom manifest while exactly cancelling the gauge variant fields with the Fadeev-Popov ghosts. In particular we are able to observe the topological stasis of the conformal mode. In section~\ref{secIRcutoff} we consider the form of the IR regulator and revisit the convexity condition \eq{ConvexGamma} for the regulated theory. Here we show how poles in the propagator can be avoided leading to a well behaved low energy limit provided the curvature satisfies $R>4\Lambda$. In light of this we employ an approximation scheme in section~\ref{secHeatKernel} whereby the early time heat kernel expansion is truncated rather than expanding in powers of the curvature. This allows us to close the Einstein-Hilbert approximation while not expanding around vanishing $R$. In the next three sections we present our results coming from these considerations while the explicit form of the flow equation is given in  appendix~\ref{App1}. The beta functions for $G_N$ and $\Lambda$ are studied in section~\ref{secBetas} and the existence of a UV attractive fixed point is shown. Then in section~\ref{secFlows} we show how the renormalisation group flow possesses asymptotically safe trajectories with a classical limit for positive, negative and vanishing cosmological constant. We then turn to the conformally reduced theory in section~\ref{secConf} where only the conformal fluctuations are quantised and their topological nature is preserved. There we find a UV fixed point which predicts the vanishing of the cosmological constant $\Lambda = 0$ on all scales. We end in section~\ref{conclusions} with with a summary of our results and our conclusions.

\section{RG for gravity and asymptotic safety} \label{secRG}

Since perturbative methods fail to give a renormalisable theory of quantum gravity, or shed light on the cosmological constant problem, one can resort to non-perturbative methods. An indispensable tool for understanding non-perturbative physics is offered by the exact (or functional)  renormalisation group  \cite{ Wilson:1973jj, RevModPhys.47.773} (for reviews see \cite{Berges:2000ew, Polonyi:2001se, Pawlowski:2005xe, Gies:2006wv, Rosten:2010vm}).  Within this framework a perturbatively non-renormalisable field theory may still be renormalised at an asymptotically safe fixed point under RG transformations. At its root is the observation that couplings of the theory, such as $G_N$ and $\Lambda$, are not constants in the quantum theory but generally depend on the momentum scale at which they are evaluated. If at high energies they tend towards an asymptotically safe fixed point their low energy values can be determined by following their RG flow into the infra-red (IR). Given such a fixed point in gravity we can then follow the flow of $G_N \cdot \Lambda$ to determine its observable value. To be a consistent theory of quantum gravity the low energy couplings must reproduce classical general relativity (plus corrections at high curvatures). Trajectories of the RG that fulfil asymptotic safety and give rise to a meaningful low energy limit can be said to be `globally safe'.

There now exists are large amount of evidence for asymptotic safety in four dimensional gravity coming from functional RG calculations \cite{Souma:1999at,  Reuter:2001ag, Lauscher:2001ya, Litim:2003vp, Codello:2008vh, Benedetti:2009rx, Ohta:2013uca, Falls:2013bv} (for reviews see \cite{Litim:2006dx,Niedermaier:2006wt,Niedermaier:2006ns,Percacci:2007sz,Litim:2008tt,Reuter:2007rv,Percacci:2011fr,Reuter:2012id}) and complimented by lattice \cite{Hamber:1999nu, Hamber:2004ew, Hamber:2009mt,Ambjorn:2012jv, Ambjorn:2013tki} and perturbative calculations \cite{Niedermaier:2009zz,Niedermaier:2010zz}. Within the functional RG approach early work concentrated on simple approximations whereby only an action of the Einstein-Hilbert form was considered \cite{Souma:1999at,  Reuter:2001ag, Lauscher:2001ya, Litim:2003vp}. Later studies have gone beyond this by including higher curvature terms \cite{Codello:2008vh, Benedetti:2009rx, Niedermaier:2010zz, Falls:2013bv, Ohta:2013uca}, general actions of the $f(R)$ type \cite{Codello:2007bd, Machado:2007ea, Falls:2013bv} and the effects of matter \cite{Percacci:2002ie, Percacci:2003jz, Vacca:2010mj, Eichhorn:2011pc, Dona:2013qba}. 

More recently more sophisticated calculations have been performed by including additional terms in the action which have a non-trivial background field dependence \cite{Manrique:2010am, Codello:2013fpa, Christiansen:2014raa, Becker:2014qya}. The nature of these non-covariant terms are in principle constrained by (modified) BRST invariance \cite{Reuter:1996cp}. At leading order these take the form of the bare gauge fixing and ghost terms arising from the Faddeev-Popov method. Beyond this approximation new terms should arise which depend on the explicit form of gauge fixing as well as the RG scheme.  In \cite{Donkin:2012ud} the background field dependence of such terms has been evaluated  via the Nielsen identities for the geometric effective action.  Although in other works the modified BRST invariance of such approximations has not been determined, the flow of covariance breaking couplings such as mass parameters \cite{Christiansen:2014raa}, wave function renormalisation \cite{ Codello:2013fpa, Christiansen:2014raa} and purely background field couplings \cite{Manrique:2010am,  Becker:2014qya} has been assessed, while  in \cite{Christiansen:2014raa} the flow of the full momentum dependent graviton propagator was evaluated. Additionally, the scale dependence of the ghost sector has been studied in \cite{Eichhorn:2009ah, Eichhorn:2010tb, Groh:2010ta}.  In each case a UV fixed point compatible with asymptotic safety has been found.

In addition to an asymptotically safe fixed point there is evidence of a non-trivial IR fixed point in quantum gravity \cite{Nagy:2012rn, Donkin:2012ud, Litim:2012vz, Rechenberger:2012pm, Christiansen:2012rx,  Christiansen:2014raa}. While earlier work suggested that this fixed point led to a non-classical running of cosmological constant, in  \cite{Christiansen:2014raa} it was found that this fixed point is for the unphysical mass parameter and that gravity behaves classically at this fixed point. Thus the existence of trajectories connecting the UV and IR fixed points imply that gravity is well defined on all length scales.

Here we will be studying the flow of the effective average action $\Gamma_k$ where $k$ denotes the RG scale down to which quantum fluctuations have been integrated out in the path integral unsuppressed. This `flowing' action obeys the exact functional renormalisation group equation \cite{Wetterich:1992yh}
\beq \label{floweq}
\partial_t \Gamma_k[\phi;\bar{\phi}] = \frac{1}{2} {\rm STr} \frac{\partial_t \mathcal{R}_k[\bar{\phi}]}{\Gamma_k^{(2)}[\phi;\bar{\phi}] + \mathcal{R}_k[\bar{\phi}]} \,,
\eeq 
obtained by taking a derivative of the action with respect to the RG time $t = \log k/k_0$. In the context of quantum gravity \cite{Reuter:1996cp} this equation has been the main tool of investigations into asymptotically safe gravity mentioned above.  In general $\Gamma_k$ depends on both the dynamical fields $\phi=\langle \varphi \rangle_k$, which are $k$ dependent averages of the fundamental fields $\varphi$ (in the presence of a source), and the non-dynamical background fields $\bar{\phi}$. The right hand side is a super-trace involving the second functional derivative $\Gamma_k^{(2)}[\phi,\bar{\phi}]$ of the action at fixed $\bar{\phi}$. The important ingredient entering \eq{floweq} is regulator function or cutoff  $\mathcal{R}_k[\bar{\phi}]$  which vanishes for high momentum modes $p^2/k^2 \to \infty$ while behaving as a momentum dependent mass term for low modes. Its presence in the denominator of the trace regulates the IR modes. Furthermore the appearance of $\partial_t \mathcal{R}_k[\bar{\phi}]$ in the numerator means the trace is also regulated in the UV due to the vanishing of the regulator for high momentum. By construction the flowing action $\Gamma_k$ interpolates between the bare action $S$ in the limit $k \to \infty$ and the full effective action $\Gamma$ when the regulator is removed at $k=0$. While the action $\Gamma_k$ need not be convex, the sum of the action and the regulator term is obtained from a Legendre transform of the regulated functional integral. This implies that the regulated inverse propagator be positive definite
\beq \label{convex}
\Gamma_k^{(2)}[\phi;\bar{\phi}] + \mathcal{R}_k[\bar{\phi}] > 0 \,,
\eeq
for all physical momentum modes included in the super-trace. Thus \eq{convex} generalises \eq{ConvexGamma} in the presence of an IR regulator. In \cite{Litim:2006nn} it was shown how convexity of the effective action follows from the flow equation \eq{floweq} for scalar fields. Furthermore, in \cite{marchais2013infrared} it was shown that convexity arises as an IR fixed point in phases with spontaneous symmetry breaking.   

In this paper we work in the Einstein-Hilbert approximation studying the flowing Euclidean action 
\bea \label{action}
&&\Gamma_k[g_{\mu\nu},...;\bar{g}_{\mu\nu}] =  \int d^4 x \sqrt{\det g_{\mu\nu}} \,  \frac{1}{16 \pi G_k}(2\Lambda_k - R(g_{\mu\nu})  + ... \,,
\eea
corresponding to general relativity with $k$ dependent couplings $G_k$ and $\Lambda_k$. The ellipses denote the extra fields and action terms coming from the gauge fixing prescription which we specify in the next section. Here we assume the conformal mode $\sigma$ has been Wick rotated from the Lorentzian action as derived from the functional measure \cite{Mazur:1989by} which ensures that the action is bounded from below. This action depends on two metrics, the dynamical metric $g_{\mu\nu}$, and the non-dynamical background metric $\bar{g}_{\mu\nu}$. The background metric is needed both to regulate the theory and to implement the gauge fixing. Once we have inserted this action into the flow equation we shall identify $\bar{g}_{\mu\nu} = g_{\mu\nu}$ in order to determine the beta functions for the flowing couplings $G_k$ and $\Lambda_k$. For a discussion of background field flows in the functional RG see \cite{Litim:2002hj}. For later convenience we also identify the wave function renormalisation of the metric $g_{\mu\nu}$ and the corresponding anomalous dimension 
\beq \label{wavefunction}
Z_k \equiv \frac{G_N}{G_k}\,,\,\,\,\,\,\,\,\,\,\,\,\,\, \eta \equiv \partial_t \ln Z_k\,,
\eeq
where $G_N$ is a constant which can be identified with the the low energy Newton's constant $G_N = G_0$ for trajectories with a classical limit.
 From the beta functions we will look for RG trajectories which emanate from a UV fixed point $G_k \to k^{-2} g_*$ and $\Lambda_k \to k^2 \lambda_*$ at high energies $k \to \infty$, while recovering classical $k$-independent couplings $G_0 =G_N$ and $\Lambda_0 = \Lambda$ when the regulator is removed in the limit $k \to 0$. Such globally safe trajectories suggest gravity is a well defined quantum field theory on all length scales. 
 
 At a non-gaussian fixed point where $g_*$ and $\lambda_*$ are finite the scaling is determined from the critical exponents $\theta_n$. These exponents appear in the linear expansion
 \beq
 \lambda^i - \lambda_*^i = \sum_n \mathcal{C}_n V^i_{n} e^{-t \theta_n} \,,
  \eeq
where $\lambda^i$ is a basis of dimensionless couplings e.g $\lambda^i = \{g,\lambda\} =\{k^2 G_k,k^{-2}\Lambda_k\}$ and the range of $n$ is equal to the range of $i$. Here $V_n^i$ are the eigen-directions and $\mathcal{C}_n$ are constants.
The exponents $-\theta_n$ (note the minus sign) and the vectors  $V_n^i$ correspond to the eigenvalues and eigenvectors of the stability matrix
\beq
M^i\,_{j}  = \left. \frac{\partial \beta^i}{\partial \lambda^j}\right|_{\lambda^i = \lambda_*^i} \,,
\eeq
where $\beta^i = \partial_t \lambda^i$ are the beta functions which vanish for $\lambda^i = \lambda_*^i$.
If $\theta_n$ is positive it corresponds to a relevant (UV attractive) direction and supports renormalisable trajectories. For negative $\theta_n$ the direction is irrelevant and $\mathcal{C}_n$ must be set to zero in order to renormalise the theory at the fixed point. Including more couplings in the approximation would introduce more directions in theory space. The criteria of asymptotic safety is that the number of relevant directions should be finite at such a UV fixed point \cite{Weinberg:1980gg}. The fewer number of relevant directions the more predictive the theory defined at the fixed point will be. High order polynomial expansions in $R$ suggest there are just three relevant directions \cite{Codello:2007bd, Machado:2007ea, Falls:2013bv} while a general argument for $f(R)$ theories imply that there is a finite number of relevant directions \cite{Benedetti:2013jk}.

\section{Physical degrees of freedom} \label{dof}

General relativity has just two massless propagating degrees corresponding to the two polarisations of the 
graviton. On the other hand conformal fluctuations, which are non-dynamical in the classical theory, are expected to play an important r\^{o}le once the theory is quantised. Our general philosophy in this paper will be to make the nature of these degrees of freedom as manifest as possible at the level of the flow equation \eq{floweq}. In this way we intended to optimise the Einstein-Hilbert approximation \eq{action} to the physics which it contains.

 In the covariant path integral quantisation, via the Faddeev-Popov prescription, the counting of propagating 
degrees of freedom comes from the ten components of the metric $g_{\mu\nu}$ minus the eight real 
degrees of freedom of the ghosts $C_\mu$ and  $\bar{C}_\mu$, each of which counts once since the action 
is second order in derivatives (i.e. the propagator will have a single pole for each independent field variable). For $d$ dimensions this gives $d(d+1)/2 - 2d = d(d-3)/2$ propagating degrees of freedom. An alternative prescription  \cite{Mazur:1989by} is to directly factor 
out of the path integral the four degrees of freedom of $g_{\mu\nu}$ corresponding to the volume of the 
diffeomorphism group
\beq
g_{\mu\nu} \to g_{\mu\nu} + \nabla_\mu \epsilon_\nu + \nabla_\nu \epsilon_\mu \,,
\eeq
which removes four unphysical degrees of freedom.
Following this procedure avoids the inclusion of ghosts in the semi-classical approximation. Instead the necessary field redefinitions  leave behind a non-trivial Jacobian in the measure of the path integral 
corresponding to a further four negative degrees of freedom. Three of these (negative) degrees of freedom correspond to a transverse vector 
which remove the three additional degrees of freedom of the transverse-traceless fluctuations of the metric $h^\upvdash_{\mu\nu}$
while an additional (negative) scalar degree of freedom cancels the conformal mode $\sigma$ in the semi-classical approximation with  $R = 4 \Lambda$ \cite{Mazur:1989by}.

 To make these cancelations visible in the flow equation \eq{floweq} we will introduce the ghosts in such a 
way that they exactly cancel the gauge fixed degrees of freedom when evaluating the flow equation for $\bar{g}_{\mu\nu}= g_{\mu\nu}$ and $C_\mu=0= \bar{C}_\nu$ \cite{Benedetti:2011ct}. This then leaves just the auxiliary degrees of freedom coming from the Jacobian plus the gauge invariant physical degrees of freedom. For simplicity we will take the metric to be that of a four sphere which is sufficient to obtain the beta functions in the Einstein-Hilbert approximation.

To this end we employ the transverse-traceless (TT) decomposition of the metric fluctuation $h_{\mu\nu} \equiv \delta g_{\mu\nu}$  given by \cite{York:1973ia}
\bea \label{TT}
h_{\mu\nu} &= &h^\upvdash_{\mu\nu} + h  \frac{1}{d} g_{\mu\nu}+ \nabla_{\nu} \xi_{\mu} + \nabla_{\mu} \xi_{\nu} + \nabla_{\mu}\nabla_{\nu}  \psi - \frac{1}{d} g_{\mu\nu} \nabla^2   \psi\,,\\[2ex]
&& \,\,\, h^\upvdash_{\mu}\,^{\mu}= 0\,, \,\,\,\,\,\,\,\,\,\,\,\,\nabla_\mu h^\upvdash_\nu\,^\mu = 0 \,,\,\,\,\,\,\,\,\,\,\,\,\nabla_{\mu}\xi^{\mu} = 0 \,. \nonumber
\eea
 Here $h^\upvdash_{\mu\nu}$ is the transverse-traceless fluctuation and $\xi_\mu$ is a transverse vector. These differential constraints have the advantage of simplifying the differential operators entering the flow equation and facilitate its evaluation. Here the spacetime dimension is taken to be $d=4$, however, there is an obvious generalisation to arbitrary dimension. In addition to the TT decomposition we re-define the trace $h= h^\mu_\mu$ in terms of the (linear) conformal mode,  
 \beq \label{sigma}
\sigma = h - \nabla^2 \psi \,,
 \eeq
which along with $h_{\mu\nu}^\upvdash$ constitute the physical degrees of freedom. 

Of course the parameterisation of the physical degrees of freedom depends on the gauge. Here we choose the gauge corresponding to $S_{\rm gf} = \frac{1}{2\alpha} \int d^dx F_\mu F^\mu$ where $F_\mu = \nabla_\lambda h^\lambda_\mu - \frac{1}{d} \nabla_\mu h^\lambda_\lambda$ and take Landau limit $\alpha \to 0$.  In this gauge contributions to the flow equation from $\xi$ and $\psi$ will just come from the gauge fixing action $S_{\rm gf}$ where the physical fields $\sigma$ and $h^\upvdash$ are absent. The gauge variant fields $\{\xi, \psi\}$ are fourth order in derivatives due to the field re-definitions ($\psi$ is momentarily sixth order but this shall be rectified shortly). In order that these contributions cancel exactly with the ghosts  we also make the ghost sector fourth order by writing $\det M= (\det M^2)^{\frac{1}{2}} $ before exponentiating the determinant of the  Faddeev-Popov  operator $M$ \cite{Benedetti:2011ct}. This introduces a third real commuting ghost $B_\mu$ as well as the anti-commuting ghosts $C_\mu$ and $\bar{C}_\mu$.
We then perform the transverse decomposition of the ghosts and an additional field redefinitions of all the longitudinal modes $\psi_L \equiv \{\psi, B,C ,\bar{C} \}$
\bea \label{ghostsdep}
C_\mu = C^T_\mu + \nabla_\mu C\,, \,\,\,\,\,\,\,\,\, \bar{C}_\mu = \bar{C}^T_\mu + \nabla_\mu \bar{C}\,,\,\,\,\,\,\, B_\mu = B^T_\mu + \nabla_\mu B\,,  \,\,\,\,\, \psi_L \to \frac{1}{\sqrt{-\nabla^2}}\psi_L \,.
\eea
This procedure leads to the Jacobians
\beq \label{Jacobian}
J_0 = ({\det}''(\Delta_0))^{\frac{1}{2}}\,, \,\,\,\,\,\, J_1 = ({\det}'(\Delta_1))^{\frac{1}{2}},
\eeq
arising from the functional measure of $\psi$ and $\xi$. 
They are determinants of the differential operators $\Delta_0 = -\nabla^2 - \frac{R}{d-1}$ and $\Delta_1 = -\nabla^2 - \frac{R}{d}$ acting on scalars and transverse vectors respectivly. The rescaling of the longitudinal modes \eq{ghostsdep} ensures that there is no Jacobian from the ghost sector and that $J_0$ is only second order in derivatives. The primes  in \eq{Jacobian} indicate that the lowest modes of $\Delta_i$ should be removed from the determinant corresponding to the negative mode and zero mode of $\Delta_0$ and the zero mode of $\Delta_1$. They are removed since the corresponding modes of $\psi$ and $\xi_\mu$ do not contribute to the physical metric fluctuations $h_{\mu\nu}$. Exponentiating the determinants in terms of auxiliary transverse fields $j_1^\mu = \{ c^\mu, \bar{c}^\mu , \phi^\mu \}$ and scalars $j_0 = \{ c, \bar{c} , \phi \}$ (where $\{c^\mu, \bar{c}^\mu,c, \bar{c}\}$ are anti-commuting) will give the four negative degrees of freedom in addition to the six degrees of freedom $h_{\mu\nu}^\upvdash$ and $\sigma$. The total bare action then reads
\beq \label{bare}
16 \pi G_N\,  S = S_{\rm EH} + S_{\rm gf} + S_{\rm gh} + \int d^4x \sqrt{\det g_{\mu\nu}}( j_0 \Delta_0 j_0  + j_{1\mu}\Delta_1 j_1^\mu )  \,.
\eeq

In the semi-classical approximation to the functional integral the integration over $\xi$ and $\psi$ will be exactly cancelled by the ghosts. In turn the conformal mode integration $\sigma$ will be cancelled by the Jacobian $J_0$ on-shell leaving only the negative mode $\sigma_-$ of $\Delta_0$.
To see these cancellations at the level of the flow equation \eq{floweq} we define the differential operator 
\beq \label{DeltaDef}
\Delta \equiv 16 \pi G_k \, \Gamma_k^{(2)} \,,
\eeq
which takes the form $\Delta = 16 \pi G_N \, S^{(2)}$  with the replacement $\Lambda \to \Lambda_k$ where $S^{(2)}$ is the second variation of the bare action \eq{bare} after a Wick rotation of the conformal mode $\sigma$. Note that due to our field redefinitions $\Delta$ is a matrix in field space.   We will normalise the fields such that all components of $\Delta$ have the form $\Delta = -\nabla^2 + ...$ ( or $\Delta = (-\nabla^2)^2 + ...$ for the fourth order parts) in order to simplify formulas. Each transverse vectors $\xi_T \equiv \{ \xi_\mu, B^T_\mu  ,C^T_\mu,\bar{C}^T_\mu\}$ and each longitudinal mode $\psi_L$  have the equal components of $\Delta$ given by the fourth order differential operators
\beq
\Delta_T = \Delta_1^2 \,, \,\,\,\,\,\,\, \Delta_L = \Delta_0^2 \,,
\eeq
however under the super-trace the corresponding terms will exactly cancel in the background field approximation. This seen by observing that in both $\xi_T$ and $\psi_L$ there are an equal number of commuting and anti-commuting fields. The remaining components of $\Delta$ are given by
\bea \label{Delta}
\Delta_\upvdash &=& \Delta_2 + 2 \left(\frac{R}{4} - \Lambda_k\right) \,, \nonumber \\
\Delta_\sigma & =& \Delta_0  + \frac{4}{3}\left(\frac{R}{4} - \Lambda_k\right) \,,\\
\Delta_0 &=& -\nabla^2 - \frac{R}{3}\,,\,\,\,\,\,\, \Delta_1 = -\nabla^2 - \frac{R}{4} \nonumber \,,
\eea   
where $\Delta_2=-\nabla^2 + \frac{R}{6}$ is the Lichnerowicz Laplacian and we have set $d=4$.
Here the conformal mode has been Wick rotated $\sigma \to i\sigma $ for all modes $ \Delta_0 \geq 0$ as derived from the functional measure \cite{Mazur:1989by}. On the other hand negative modes $\sigma_-$ of this operator should be wick rotated trivially  \cite{Mazur:1989ch}. On the sphere there is just one such mode corresponding to the constant mode which gives an eigenvalue of the operator $-\Delta_\sigma$ of $a_{-} = + \frac{R}{3} - \frac{4}{3}\left(\frac{R}{4} - \Lambda_k\right)$. Physically this mode corresponds to a rescaling of the radius of the four sphere \cite{Mazur:1989ch}.  Taking into account all contributions and the cancellation of the ghost and gauged fixed parts the flow equation reads
\bea \label{flowhere}
\partial_t \Gamma_k &=& \sum_i \mathcal{S}_{i} \equiv \frac{1}{2} \Tr\left[ \frac{\partial_t\mathcal{R}_{\upvdash,k}}{Z_k \Delta_\upvdash+ \mathcal{R}_{\upvdash,k} }\right] + \frac{1}{2} \Tr''\left[ \frac{\partial_t\mathcal{R}_{\sigma,k}}{Z_k \Delta_\sigma + \mathcal{R}_{\sigma,k}}\right] + \frac{1}{2} \left[ \frac{\partial_t\mathcal{R}_{-,k}}{Z_k a_- + \mathcal{R}_{-,k}}\right]  \nonumber \\
&-& \frac{1}{2}\Tr''\left[ \frac{\partial_t\mathcal{R}_{0,k}}{Z_k \Delta_0 + \mathcal{R}_{0,k}}\right]- \frac{1}{2}\Tr'\left[ \frac{\partial_t\mathcal{R}_{1,k}}{Z_k \Delta_1 + \mathcal{R}_{1,k}}\right] \, ,
\eea
where $\mathcal{S}_i$ are the various traces $i =\{\upvdash,\sigma,-,0,1\}$ and the prime indicates the excluded modes. We observe that by going on-shell $\Lambda_k = R/4$ we have $\Delta_\sigma = \Delta_0$ indicating that the conformal fluctuations are removed by those of $j_0$ arising from the scalar Jacobian \eq{Jacobian}. The traverse vector fluctuations should then remove the three non-propagating degrees of freedom of $h_{\mu\nu}^\upvdash$.

 Since the on-shell condition is not generally satisfied along the flow these cancellations do not occur exactly.
 However, the above reasoning implies a natural pairing of the contributions $\mathcal{S}_{\rm grav}\equiv\mathcal{S}_{2} + \mathcal{S}_{1}$ and $\mathcal{S}_{\rm conf} \equiv \mathcal{S}_{\sigma} + \mathcal{S}_0$ which carry two and zero propagating degrees of freedom respectively. These contributions are then identified with physical graviton and conformal fluctuations of spacetime.  A standard approximation scheme to test asymptotic safety is to only quantise the conformal mode $\sigma$. At the level of \eq{flowhere}  this could be achieved in two ways. On one hand we could make this approximation by only including $\mathcal{S}_\sigma$. On the other hand this would mean $\sigma$ is a propagating degree of freedom since the Jacobian contribution is not there to cancel its on-shell dynamics \footnote{In $f(R)$ gravity the conformal mode becomes fourth order and is a propagating degrees of freedom, however not including $\mathcal{S}_{0}$ would then mean we have two propagating scalars.}. This suggests that a more consistent approximation is achieved by keeping both contributions to $\mathcal{S}_{\rm conf}$. We will come back to this point in section~\ref{secConf} where we consider these approximations.

\section{Infra-red cutoff and the cosmological constant} \label{secIRcutoff}
We now turn to the form of the IR regulator $\mathcal{R}_k$ which must be specified in order to evaluate the traces in \eq{flowhere}. We will take particular care to regulate modes in such a way that the convexity condition \eq{convex} is satisfied. This point has been stressed \cite{Benedetti:2013jk} in the context of the $f(R)$ approximation to asymptotic safety and was discussed in \cite{Folkerts:2011jz} for Yang-Mills coupled to gravity. We note that $\mathcal{R}_k$ depends on the background field which translates to a dependence on the scalar curvature $R$. As we shall see this suggests a specific form of the regulator depending on $R$ and the scale dependent cosmological constant $\Lambda_k$.
 In general the form of the regulator will be
\beq
\mathcal{R}_k = \frac{1}{16 \pi G_k} R_k(z)\,,
\eeq
where the cutoff function $R_k$ (not to be confused with the scalar curvature $R$) should vanish in the limit $k \to 0$ for all values of $z>0$. Here $z$ should be (the eigenvalue of) some differential operator of the form $z = -\nabla^2 + U$ where $U$ is some potential. In the classifications of \cite{Codello:2008vh} a cutoff for which $U=0$ is referred to as {\it type I}, whereas a curvature dependent potential $U=U(R)$ with no $k$ dependence is called a { \it type II} cutoff, finally a general $k$ dependent potential $U=U_k(R)$ is termed {\it type III}. 

In curvature expansions one expands the trace in powers of the curvature in order to extract the beta functions for the running couplings $G_k$ and $\Lambda_k$.  This may lead to poles in the propagator which can be seen by looking at the components of $ \Delta$ in \eq{Delta} for the conformal and transverse traceless fluctuations.  
Setting $R=0$ will create poles at $-\nabla^2 = 2\Lambda_k$ and $-\nabla^2 = \frac{4}{3} \Lambda_k$ in the unregulated propagator.
These are clearly artefacts of expanding in the curvature and have no obvious physical meaning. On the other hand the graviton is a massless degree of freedom and should have a pole in its propagator at zero momentum.
Indeed if we instead set the background metric to a solution of the equation of motion $R= 4 \Lambda_k$  we have  $\Delta_\upvdash =\Delta_2$ and  $\Delta_\sigma= \Delta_0$. For the regulated propagators of $\sigma$ and $h^\upvdash_{\mu\nu}$ we have potential poles at $R=0$ for
\bea \label{poles}
P_\sigma(R=0) \equiv -\nabla^2 - \frac{4}{3} \Lambda_k + R_{\sigma,k}\stackrel{!}{=} 0\,, \\
P_\upvdash(R=0) \equiv -\nabla^2 - 2 \Lambda_k +  R_{\upvdash,k} \stackrel{!}{=} 0\,.
\eea
However taking $R$ equal or greater to its on-shell value $R \geq 4 \Lambda_k$ ensures that $\Delta \geq 0$ and that no unphysical pole can be present (note that $\Delta_0$ and $\Delta_1$ are positive definite since the negative and zero modes are not excluded). Now along the flow we only require $\Gamma^{(2)} + \mathcal{R}_k> 0$ so the flowing $\Lambda_k$ need not satisfy $\Lambda_k  \leq R/4$ for all $k$. Instead we may regulate this potential pole by an appropriate choice of $\mathcal{R}_k$. On the other hand this must be done in such a way that the regulator function $\mathcal{R}_k$ vanishes in the limit $k \to 0$ such that all modes are integrated out unsuppressed.

 Now say we choose a curvature independent type III cutoff $z = -\nabla^2 - 2 \Lambda_k$ in order that we remove the poles \eq{poles} then $z$ can take negative values for eigenvalues $p^2$ of the Laplacian $-\nabla^2$ for which $p^2< 2 \Lambda_k$. For these eigenvalues the regulator would not vanish in the limit $k \to 0$. For example if we take the optimised cutoff \cite{Litim:2001up} $R_k(z)= (k^2 - z)\theta(k^2-z)$ at $k=0$ we have $R_0(z)= -z\theta(-z)$ which only vanishes if $z$ is positive and therefore not all modes will be integrated. If we instead take $z= \Delta$, given by \eq{Delta}, we can ensure that $z$ is positive at $k=0$ provided the curvature satisfies $R \geq 4 \Lambda_0$. On the other hand modes for which $z<0$ for finite $k$ can still be regulated.
Here we will therefore use a type III regulator of the form
\beq \label{RegChoice}
\mathcal{R}_k = \frac{1}{16 \pi G_k} R_k(\Delta)\,.
\eeq
This choice has been studied in \cite{Codello:2008vh} where it was shown that asymptotically safe trajectories can reach a classical limit at $k=0$ for positive $\Lambda$. Such a regulator is called a {\it spectrally adjusted} cutoff since it cuts off modes with respect to the full $k$ dependent inverse propagator $\Delta$.
We observe that the vanishing of the regulator \eq{RegChoice}  at $k=0$ for different values of the curvature $R$ coincides with the convexity condition \eq{ConvexGamma} provided $G_0>0$ . Here we will assume that $R> 4 \Lambda_0$ such that $\mathcal{R}_k$ indeed vanishes when we take the IR limit. In particular at classical infra-red fixed points for which $G_k$ and $\Lambda_k$ approach constants the condition on $R$ in Planck units then depends on the value of the dimensionless product $G_0 \cdot \Lambda_0$. We will return to this in section~\ref{secFlows} where discuss renormalisable trajectories that reach a line of such fixed points.

\section{Truncated heat kernel expansion} \label{secHeatKernel}

To compute the beta functions of $G_k$ and $\Lambda_k$ we must evaluate the traces appearing on the right side of the flow equation. However in order close our equations an approximation scheme is needed since the traces will in general lead to curvature terms not present in our original action. 
We observe that each of the traces in \eq{flowhere} are functions $f(\Delta)$ of the differential operator \eq{Delta}.  
As a first step we can express the trace in terms of the heat kernel via an anti-Laplace transform with respect to $\Delta$ and expand in the early time $s$ expansion. They then have the form
\bea \label{expansion}
\mathcal{S} = \Tr[f(\Delta )] &=& \int ds \, \Tr [e^{-\Delta s}] \,\tilde{f}(\tau)  
\approx \frac{1}{(4 \pi)^{\frac{d}{2}}}\sum_{n=0}^{\infty} Q_{\frac{d}{2}-n}[f] A_{n}(R, \Lambda_k) \,,
\eea
where we suppress the field index $i$.
Here $A_{n}(R, \Lambda_k)$ are the Seeley-DeWitt coefficients coming from the expansion of the heat kernel $H_\Delta(s) \equiv e^{-\Delta s}$  which obeys the heat equation $\Delta H_\Delta + \partial_s H =0$, subject to the initial condition $H(0)=I$ where $I$ is the identity operator. These coefficients depend on both the curvature and the scale dependent cosmological constant $\Lambda_k$.  The appearance of the cosmological constant inside the heat kernel coefficients is a direct consequence of the fact that the covariant momentum (i.e. eigenvalues of $\Delta$) explicitly depends on $\Lambda_k$. The functionals $Q_m[f] \equiv \int_0^\infty d \tau \tau^{-n} \tilde{f}(\tau)$ depend on the argument $f(z)$ of the traces given it \eq{flowhere}.  For $m >0$ they are given by the following integrals over the covariant momentum $z$,
\beq \label{Qs}
Q_m = \frac{1}{\Gamma(m)}\int^\infty_0 dz z^{m-1} f(z) \,.
\eeq
Note that these integrals are over $z \geq 0$ and therefore by adopting the heat kernel evaluation we automatically regulate modes $z<0$ in a sharp way.
This can be traced back to the anti-Laplace transform which only converges for $\Delta \geq 0$. 

Within the standard approach, where the momentum is independent of $\Lambda_k$, one would simply expand to order $R$ and neglect the higher order terms. Here we take a different approach and use the heat kernel expansion itself as the basis of our approximation scheme. That is we drop all heat kernel coefficients for $n>n_{\rm max}$ where we take $n_{\rm max}=1$. Additionally we drop the single negative conformal mode $\sigma_-$ whose contribution is proportion to $d^4x \sqrt{\det g_{\mu\nu}} R^2$. To better the approximation we can increase $n_{\rm max}$ systematically and assess the convergence properties \cite{Falls:2013bv}. Note that this differs from a curvature expansion since all higher order heat kernel coefficients will depend on terms linear in $R \Lambda_k^{n-1}$ and $\Lambda^{n}_k$  (such terms have also been neglected in \cite{Dou:1997fg} in order to be able to go on-shell by assuming $\Lambda_k$ is of order $R$). A truncation of the heat kernel expansion rather than the curvature expansion is therefore different approximation scheme which should have different convergence properties. Since it is not strictly a curvature expansion (around any point zero or otherwise) it does not necessitate that the curvature is `small' however the early time heat kernel expansion should be expected to accurately evaluate the traces in the high momentum limit $R/z \sim R/k^2 \to 0$.

Our justification for this approximation is twofold. First this keeps the cosmological constant appearing to the combination $R- 4\Lambda_k$ so as not to upset the on-shell limit. Another approach to this, put forward in \cite{Benedetti:2011ct}, is to expand the trace around $R= 4\Lambda_k$ which involves evaluating the the trace via an approximation of the spectral sum.
However, our second motivation is to get the approximation well suited to the power like divergence that renormalise $\Lambda_k$ and $G_k$. These come from the large momentum limit of the trace. Since the early time heat kernel expansion correctly evaluates these terms in the asymptotic limit, embodied in the first two heat kernel coefficients, it is ideally suited to the Einstein-Hilbert approximation. What we neglect are the logarithmic divergences which renormalise the curvature squared terms at order $n=2$ (and the IR divergent terms $n>2$). Since these are also absent in the left hand side of the flow equation this approximation is self-consistent. These corrections are then naturally included in the $n_{max} =2$ approximation where curvature squared terms are included. This approach is then in line with the bootstrap approach to asymptotic safety \cite{Falls:2013bv} without having to specify $R=0$ as an expansion point.

\section{Beta functions and UV fixed point} \label{secBetas}
We are now in the position to derive the beta functions $\beta_g = \partial_t g$ and $\beta_\lambda = \partial_t \lambda$ within the set-up outlined in the preceding sections. The vanishing of the beta functions for non-vanishing $\{g_*,\lambda_*\}$ indicate a non-gaussian fixed point where the theory may be renormalised.

\subsection{Flow equation and threshold constants}

The explicit form of the flow equation is given in the appendix~\ref{App1} where we also give the heat kernel coefficients $A_n$.
Each component of $\Delta$ in \eq{Delta} has the form $\Delta_i=-\nabla^2 + U_i$ where the potentials $U_i = U_i(R,\Lambda_k)$ (given in \eq{U}) are linear in the scalar curvature $R$ and the cosmological constant $\Lambda_k$. The corresponding heat kernel coefficients $A_{i,n} = \int d^dx \sqrt{\det g_{\mu\nu}}\, a_{i,n}$  which depend on these potentials are then given by \eq{As} in the appendix.
We also need to evaluate the $Q_n$ functions \eq{Qs} which depend on the regulator functions $R_k$ and the beta functions themselves since $\mathcal{R}_k$ depends on both $\Lambda_k$ and $G_k$. Here we will only need to evaluate $Q_m$ for $m=1,2$ where, in the sum \eq{expansion}, $Q_{2}$ appears at $n=0$ and $Q_1$ appears at $n=1$. For all $m>0$ they have the form
\beq \label{Qsexplicit}
Q_{m,i} = (-1)^{[i]} \frac{k^{2m}}{2}  \left(\Phi_m[R_k] + \tilde{\Phi}_m[R_k] \eta +  \hat{\Phi}_m[R_k]  \dot{U}_i\right) \,.
\eeq
Here the dot denotes a derivative with respect to the RG time $t=\ln (k/k_0)$. The anomalous dimension is given by $\eta \equiv \dot{Z}_k/Z_k= - \eta_N \equiv - \dot{G}_k/G_k$ (see \eq{wavefunction}) which we take to be the same for each field and takes the value $\eta_* =2$ at a non-trivial fixed point.
The $[i]$ in the exponent of $-1$ takes values $[2]=0=[\sigma]$ for the physical degrees of freedom and $[0]=1=[1]$ for the `anti'-degrees of freedom as dictated by the super-trace. The `threshold constants'  $\Phi_m$, $ \tilde{\Phi}_m$ and $ \hat{\Phi}_m $  are given by the following regulator $R_k$ dependent integrals evaluated for $k=1$,
\beq \label{Threshold}
\Phi_m = \int_0^\infty dz  z^{m-1} \frac{\dot{R}_1(z)}{z+ R_1(z)},\,\,\,\,\tilde{\Phi}_m = \int_0^\infty dz  z^{m-1} \frac{R_1(z)}{z+ R_1(z)}, \,\,\,\,\,\,\, \hat{\Phi}_m = \int_0^\infty dz  z^{m-1} \frac{R_1^{\,'}(z)}{z+ R_1(z)} \,,
\eeq 
where the prime denotes derivative with respect to the covariant momentum $z$.
 Since the threshold constants only depend on the shape function $R_k$ and are independent of the curvature and couplings they will just be numbers once the regulator is specified. We note that for any regulator function the threshold constants have a definite sign
\beq \label{thresholdsign}
\Phi_m > 0\,, \,\,\,\, \tilde{\Phi}_m > 0\,, \, {\rm and} \,\,\,\,\,\, \hat{\Phi}_m < 0 \,.
\eeq
This information allows us to determine physical fixed points and their properties without specifying the form of $R_k$.
The final form of the flow equation in terms of the dimensionless coupling $g= k^2G_k$, $\lambda = k^{-2} \Lambda_k$ and the constants \eq{Threshold} is given in \eq{flowexplicit} with \eq{JS},\eq{sigmaS} and \eq{gS}. 
 We note that the `one-loop' approximation where by $\Gamma^{(2)}_k$ is replaced by $S^{(2)}$ in the right hand side of the flow equation translates to putting $\eta=0=\dot{U}_i$. This can also be achieved by setting $\tilde{\Phi}_n=0= \hat{\Phi}_n$. We will consider this approximation in section~\ref{secOneloop}.

\subsection{Regulator functions}
Here we consider the class of exponential functions of the form
\beq \label{Rexp}
R_k^{\rm exp}(z) = k^2 \frac{1}{2 \exp\left[c \frac{z^b}{k^{2b}}\right] -1}\,,
\eeq
where $b$ is a free parameter which we study in the range $2 \leq b \leq 30$ and we set $c= \ln 3/2$. Increasing $b$ sharpens the division between low and high modes. 
In addition to the exponential regulators we also consider the optimised regulator function \cite{Litim:2001up}
\beq \label{Ropt}
R_k^{\rm opt}(z) = (k^2 - z) \theta(k^2-z)\,,
\eeq
where $\theta(x)$ is the Heaviside theta function. We use the notation $\stackrel{\rm opt}{=}$ for quantitates evaluated with \eq{Ropt}. Plugging these functions into the integrals \eq{Threshold} we obtain the numerical values for the threshold constants. For example with the optimised cutoff function we have $\Phi_1 \stackrel{\rm opt}{=}  2$,  $\Phi_2 \stackrel{\rm opt}{=}  1$, $\tilde{\Phi}_1\stackrel{\rm opt}{=} \frac1 2 $,  $\tilde{\Phi}_2\stackrel{\rm opt}{=} \frac1 6 $, $\hat{\Phi}_1\stackrel{\rm opt}{=} -1 $ and $\hat{\Phi}_2 \stackrel{\rm opt}{=}  -\frac{1}{2}$. The curvature dependence of the traces comes solely from heat kernel coefficients \eq{As}. 

\subsection{Beta functions and fixed points}

\begin{figure}[t]
\includegraphics[width=.7\hsize]{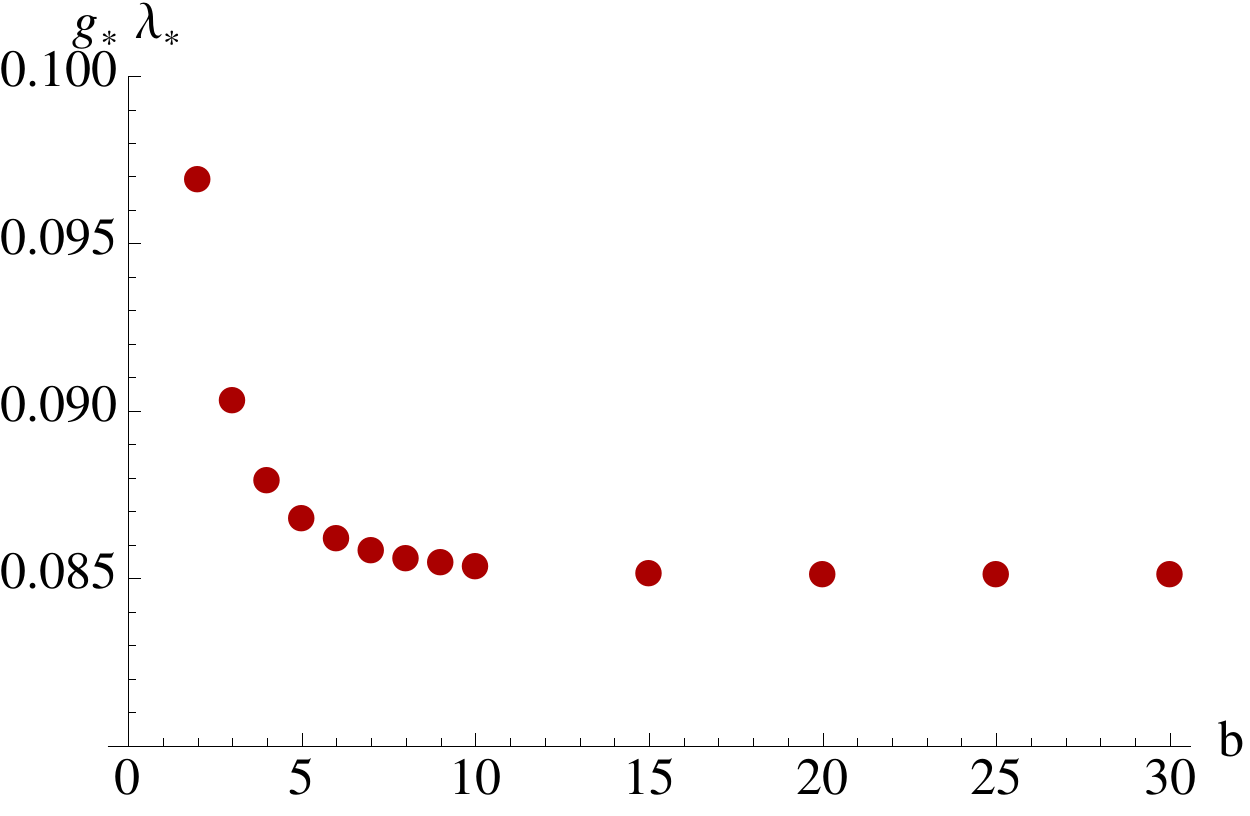}
\caption{\label{glambda}   We plot the regulator dependence of $g_* \cdot \lambda_*$ at the non-perturbative fixed point \eq{UVFP} as function of the parameter $b$ appearing in the exponential cutoff \eq{Rexp}. Increasing $b$ sharpens the cutoff of IR modes and we observe a convergence of $g_* \cdot \lambda_*$ }   
\end{figure}

Before specifying the regulator $R_k$ the beta functions $\beta_g = \partial_t g$ and $\beta_\lambda = \partial_t \lambda$ may be expressed explicitly in terms of the threshold constants \eq{Threshold} with \eq{thresholdsign},
\beq \label{betaglambda}
\beta_g=g \left(2+\frac{g \left(-438 g \Phi _2 \hat{\Phi }_1+\Phi _1 \left(752 g \lambda  \hat{\Phi }_1+99 \left(6 \pi +17 g \hat{\Phi
   }_2\right)\right)\right)}{9 \left(6 \pi +17 g \hat{\Phi }_2\right) \left(-4 \pi +11 g \tilde{\Phi }_1\right)+2 g \hat{\Phi }_1
   \left(376 g \lambda  \tilde{\Phi }_1-3 \left(50 \pi  \lambda +73 g \tilde{\Phi }_2\right)\right)}\right) \,,
\eeq
\beq
\beta_\lambda =-2 \lambda -\frac{9 g \left(2 \pi  \left(\lambda  \Phi _1+6 \Phi _2\right)+33 g \left(-\Phi _2 \tilde{\Phi }_1+\Phi _1 \tilde{\Phi
   }_2\right)\right)}{-216 \pi ^2+6 g \pi  \left(-50 \lambda  \hat{\Phi }_1-102 \hat{\Phi }_2+99 \tilde{\Phi }_1\right)+g^2 \left(752
   \lambda  \hat{\Phi }_1 \tilde{\Phi }_1+1683 \hat{\Phi }_2 \tilde{\Phi }_1-438 \hat{\Phi }_1 \tilde{\Phi }_2\right)}\,.
\eeq 
These beta-functions are evidently non-perturbative.
Solving for fixed points $\beta_g = 0 =  \beta_\lambda$ we find a gaussian fixed point $\{g=0,\, \lambda=0\}$
and a pair of non-gaussian fixed points one of which is at positive $g$ and $\lambda$ for all cutoff functions. Due to the structure of the flow equation \eq{flowexplicit} we always find exactly two non-gaussian fixed points in the complex plane for all regulators, one is at positive $g_*$ and the other at negative $g_*$. To ensure the convexity condition \eq{convex} only the fixed point for positive $g_*$ is physical. In terms of the threshold constants the physical fixed point couplings are given by
\bea \label{UVFP}
g_*&=& \frac{576 \pi }{208 \Phi _1+416 \tilde{\Phi }_1+73 \left(-17 \hat{\Phi }_2+\sqrt{\left(8 \Phi _1+17 \hat{\Phi }_2+16 \tilde{\Phi
   }_1\right){}^2-96 \hat{\Phi }_1 \left(\Phi _2+2 \tilde{\Phi }_2\right)}\right)} \nonumber\,, \\[2ex]
\lambda_*& =& \frac{8 \Phi _1+17 \hat{\Phi }_2+16 \tilde{\Phi }_1-\sqrt{\left(8 \Phi _1+17 \hat{\Phi }_2+16 \tilde{\Phi }_1\right){}^2-96 \hat{\Phi }_1
   \left(\Phi _2+2 \tilde{\Phi }_2\right)}}{32 \hat{\Phi }_1}\,.
\eea
which, due to \eq{thresholdsign},  can be seen to be both manifestly real and positive.
For the optimised cutoff we have
\bea
g_* \stackrel{\rm opt}{=} \frac{36 \left(73 \sqrt{1473}-2489\right) \pi }{51703} \approx 0.68405,\,\,\,\,   \lambda_* \stackrel{\rm opt}{=} \frac{1}{64} \left(\sqrt{1473}-31\right) \approx 0.115308\,.
\eea 
These quantities are not universal and may have a strong regulator dependence. On the other hand the dimensionless product $G_k^*\cdot \Lambda_k^* = g_*\cdot\lambda_*$ is expected to be universal.
For the optimised regulator function \eq{Ropt} the product is given by
\beq
 g_*\cdot\lambda_* \approx 0.0788761\,.
\eeq
In fig.~\ref{glambda} we plot the dependence of $g_* \cdot \lambda_*$ on the regulator parameter $b$ for the regulator function \eq{Rexp}. As $b$ is increased we see a convergence.

\begin{figure}[t]
\centering
\begin{center}
\unitlength0.001\hsize
\begin{picture}(1000,470)
\put(10,0){\includegraphics[width=.47\hsize]{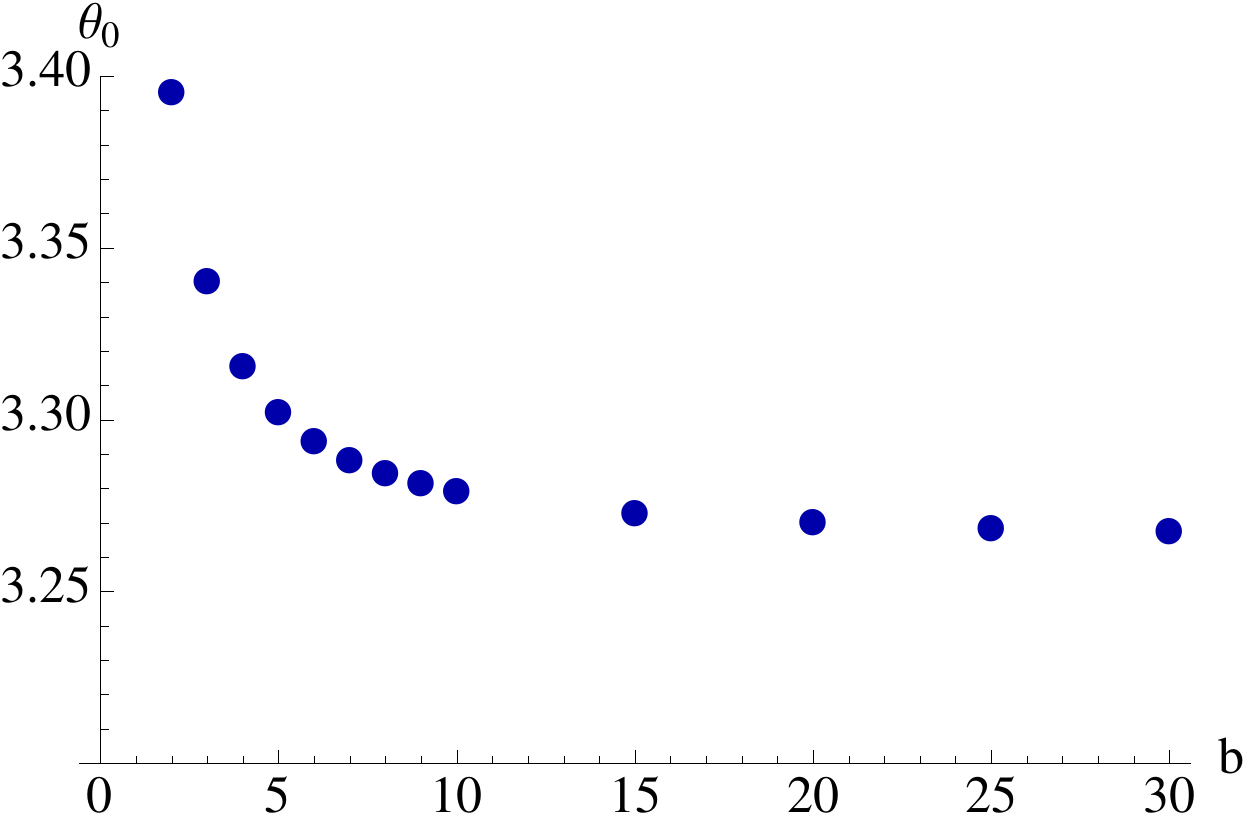}}
\put(500,0){\includegraphics[width=.48\hsize]{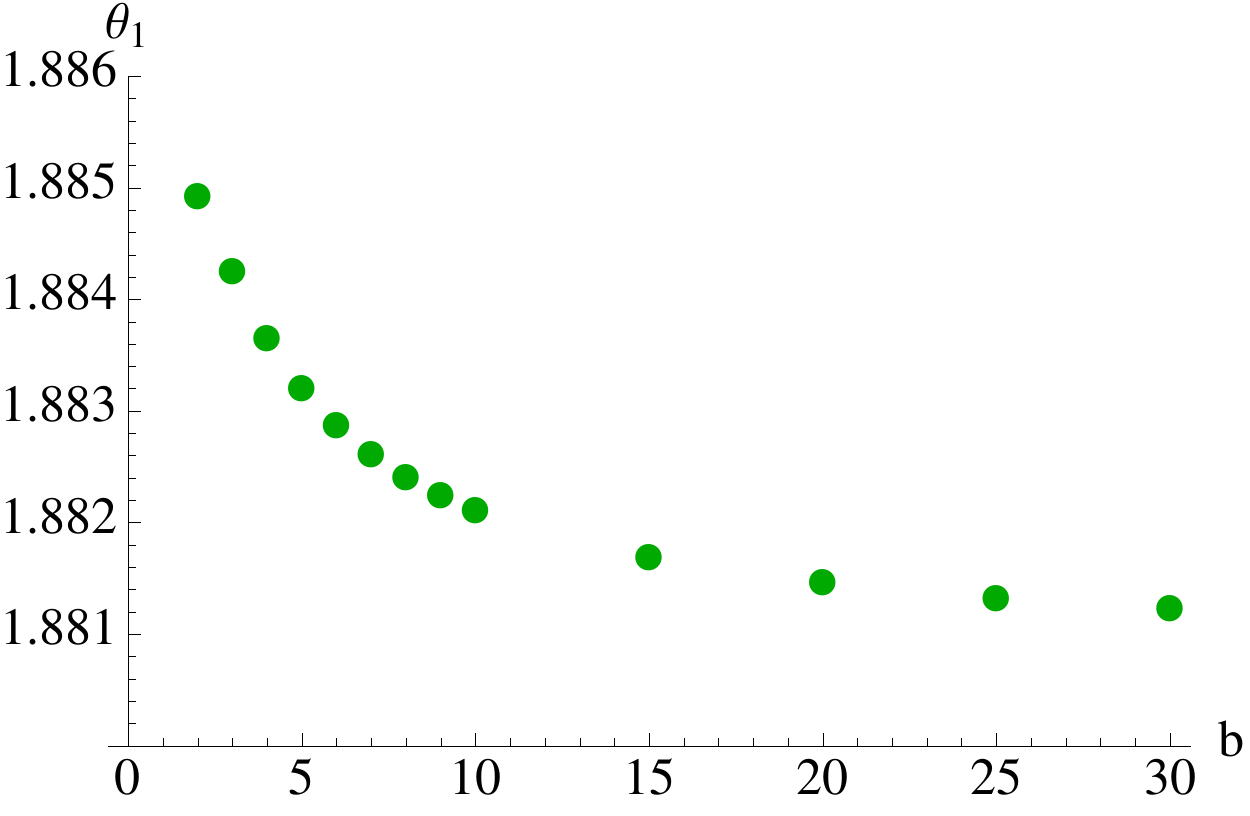}}
\end{picture}
\caption{\label{theta} We plot the UV attractive critical exponents at the non-perturbative fixed point \eq{UVFP} evaluated using the exponential cutoff \eq{Rexp}. As the sharpening parameter $b$ is increased we see a convergence of both critical exponents. These values should also be compared to those obtained with the optimised cutoff \eq{Ropt} given by \eq{thetaopt}.    }
\end{center}
\end{figure}

Expressions for the critical exponents can also be obtained in terms of the threshold constants but they lengthy so we do not include them here. For all regulators considered they are each both real and relevant.
Using the optimised cutoff  \eq{Ropt} the critical exponents are given by 
\beq \label{thetaopt}
\theta_0 \stackrel{\rm opt}{\approx} 3.35126 \,, \,\,\,\,\,\, \theta_1 \stackrel{\rm opt}{\approx} 1.87582\,.
\eeq  
In fig.~\ref{theta} we plots the dependence of the critical exponents on $b$ for the exponential cutoff functions \eq{Rexp}. We note that they are close to the values \eq{thetaopt} and converge as $b$ is increased. Here we use the convention that the more relevant critical exponent is denoted $\theta_0$.

 Numerically the critical exponents calculated with the optimised cutoff function \eq{Ropt} are within $\approx 16\%$ and $\approx 6\%$ of the gaussian critical exponents $\theta_{\rm G, 0} = 4$ and $\theta_{\rm G 1} =2$ consistent with the bootstrap approach put forward in \cite{Falls:2013bv}. However it is also instructive to look at the corresponding eigenvectors. These, unlike the critical exponents, depend on the parameterisation of the fixed point coordinates. Since the (non-perturbative) power counting comes from the canonical dimension of the operators in the action \eq{action} it therefore makes sense to consider the running  vacuum energy $ \rho_k = \Lambda_k/G_k$ and the running Planck mass (squared)  $M_{k}^2= G_k^{-1}$ which appear as the coefficients of these operators. In this basis the eigenvectors are given by
\beq
\mathbf{V}_0 \equiv \{V^{\rho}_0, V^{M^2}_0\}\approx\{0.37688, 0.926262\} \,, \,\,\,\,\,\,\,\,\,\mathbf{V}_1 \equiv \{V^{\rho}_1, V^{M^2}_1\}\approx\{0.987898, 0.155106\} \,.
\eeq
for the optimised cutoff.
Interestingly we observe that the more relevant eigenvector $\mathbf{V}_0$ points more strongly in the direction of $M_{k}^2$ rather than the vacuum energy $\rho_k$ direction and vice versa for $\mathbf{V}_1$. This indicates that $M_{k}^2$ becomes more relevant in the UV and $\rho_k$ less relevant. With the exponential cutoff \eq{Rexp},  less relevant eigenvector also points more strongly in the $\rho_k$ direction.

It is intriguing to note that we obtain real critical exponents and not a complex conjugate pair found in previous Einstein-Hilbert approximations \cite{Souma:1999at,  Reuter:2001ag, Lauscher:2001ya, Litim:2003vp}, including the on-shell approach \cite{Benedetti:2011ct}.
However real exponents have been found in work that goes beyond this approximation by utilising vertex expansions around flat space \cite{Christiansen:2012rx, Codello:2013fpa, Christiansen:2014raa}. 
Also the critical exponents have been shown to be real provided a global $f(R)$-type fixed point solution exists \cite{Benedetti:2013jk}. This suggests that by not explicitly expanding in powers of the curvature we have a better approximation to such a solution.

\subsection{One-loop scheme independence} \label{secOneloop}
The semi-classical or `one-loop' \footnote{This is a slight abuse of language since the flow equation \eq{floweq} is manifestly one-loop exact. By one loop we therefore mean the semi-classical approximation, keeping quantum effects up to order $\hbar$.}  approximation  to the flow equation \eq{floweq} is achieved by putting $\Gamma_k^{(2)} = S^{(2)}$ in the right hand side. This leads to the equation
\beq \label{oneloop}
\partial_t \Gamma_k^{\rm{one-loop}}= \frac{1}{2} {\rm STr} \left[\frac{\partial_t \mathcal{R}_k}{S^{(2)} + \mathcal{R}_k}\right] \,,
\eeq
where the regulator function $\mathcal{R}_k$ should be modified accordingly.
To obtain this approximation at the level of our beta-function we neglect the running of $G_k$ and $\Lambda_k$ on the right-hand side of the flow equation
which is equivalent to putting $\tilde{\Phi}_n= 0 = \hat{\Phi}_n$. The beta-functions then simplify to the form
\bea
\beta_g &=& 2 g-\frac{11 g^2 \Phi _1}{4 \pi } \,, \\[2ex]
\beta_\lambda &=& \frac{-24 \pi  \lambda + g \lambda  \Phi _1+6 g \Phi _2}{12 \pi }\,.
\eea
These beta-functions have a single non-trivial UV fixed point
\beq
g_* = \frac{8 \pi }{11 \Phi _1}\,, \,\,\,\,\,\,\,  \lambda_* = \frac{3 \Phi _2}{16 \Phi _1} \,,
\eeq
with regulator $R_k$ {\it independent} critical exponents
\beq
\theta_ 0 = 2\,, \,\,\,\, \theta_1 = \frac{64}{33} \approx 1.939\,.
\eeq
The more relevant exponent $\theta_0 = \theta_{G,1}$ is just the canonical mass dimension of the Planck mass squared $M_{\rm Pl}^2$ whereas $\theta_1$ is a true quantum correction. In \cite{Christiansen:2014raa} a real critical exponent for $g$ of $\theta = 2$ has been found in agreement with the
one-loop result found here. 
This scheme independence can be traced to our treatment of the cosmological constant
and is directly linked to the use of the truncated heat kernel expansion suggesting that this approximation may better converge to the physical result. Ultimately this can be tested by increasing $n_{\rm max}$ in a systematic way \cite{Falls:2013bv}.

\begin{figure}[t]
\includegraphics[width=1.0\hsize]{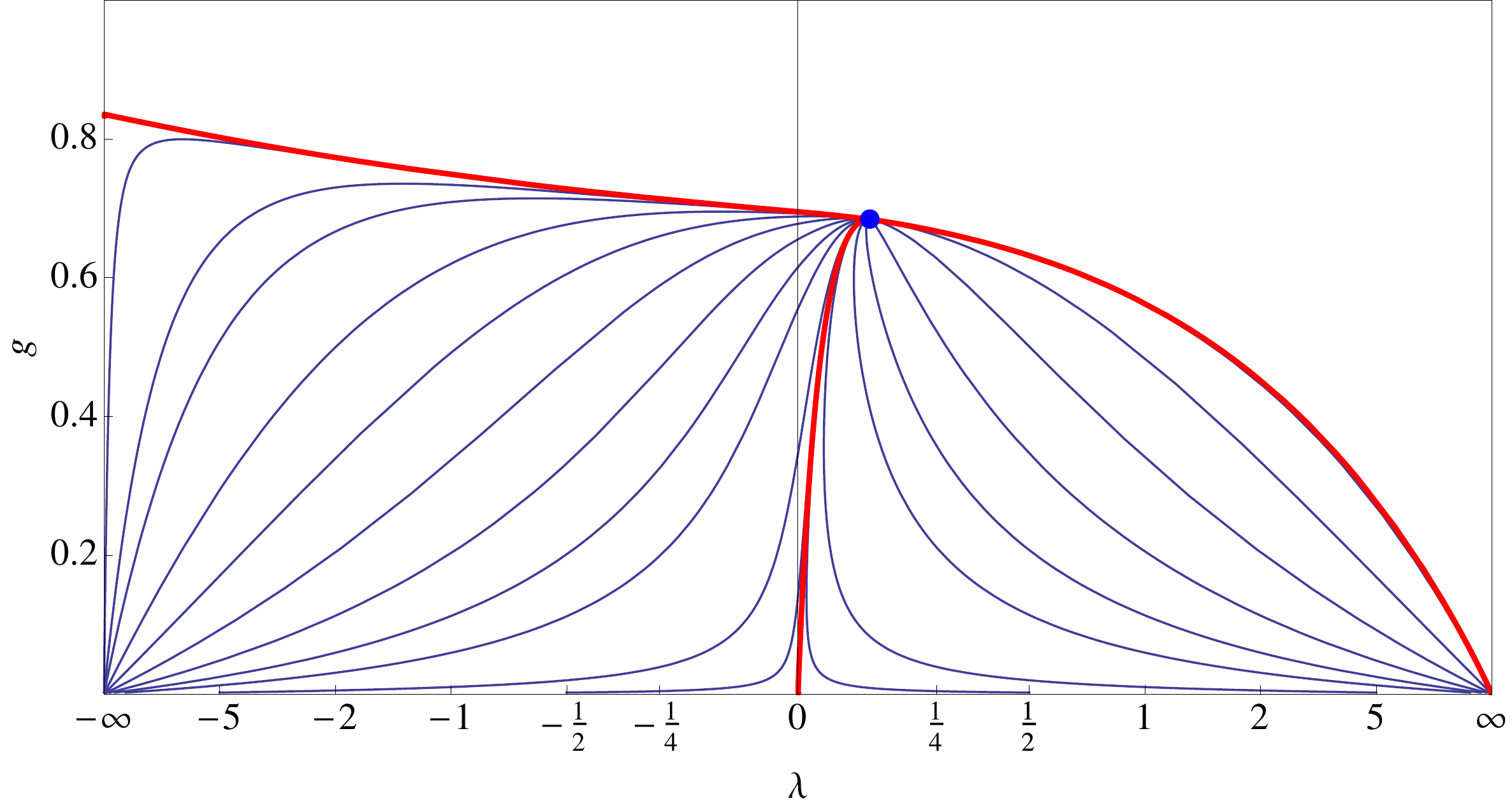}
\caption{\label{Trajectories} Phase diagram in the $\{\lambda ,g \}$ parameterisation. We plot the globally safe trajectories originating from the asymptotically safe fixed point (blue dot) for $k \to \infty$ and ending in classical general relativity for $k =0$.  Each trajectory corresponds to a different value of $G_N \cdot \Lambda$ for $k \to 0$ lying in the range $-\infty < G_N \cdot \Lambda \leq \tau_{\rm max}$. The red lines are the trajectories  $G_N \cdot \Lambda = \tau_{\rm max} = \frac{18 \pi }{25}$, $G_N \cdot \Lambda = 0$ and $G_N \cdot \Lambda = -\infty$ corresponding to the infinite fixed point $\lambda =-\infty$, $g_*= \frac{25 \pi }{94}$. In the regions where no trajectories are plotted no globally safe trajectories exist. Here we use the optimised cutoff \eq{Ropt}.  }

\end{figure}

\begin{figure}[t]
\includegraphics[width=1.0\hsize]{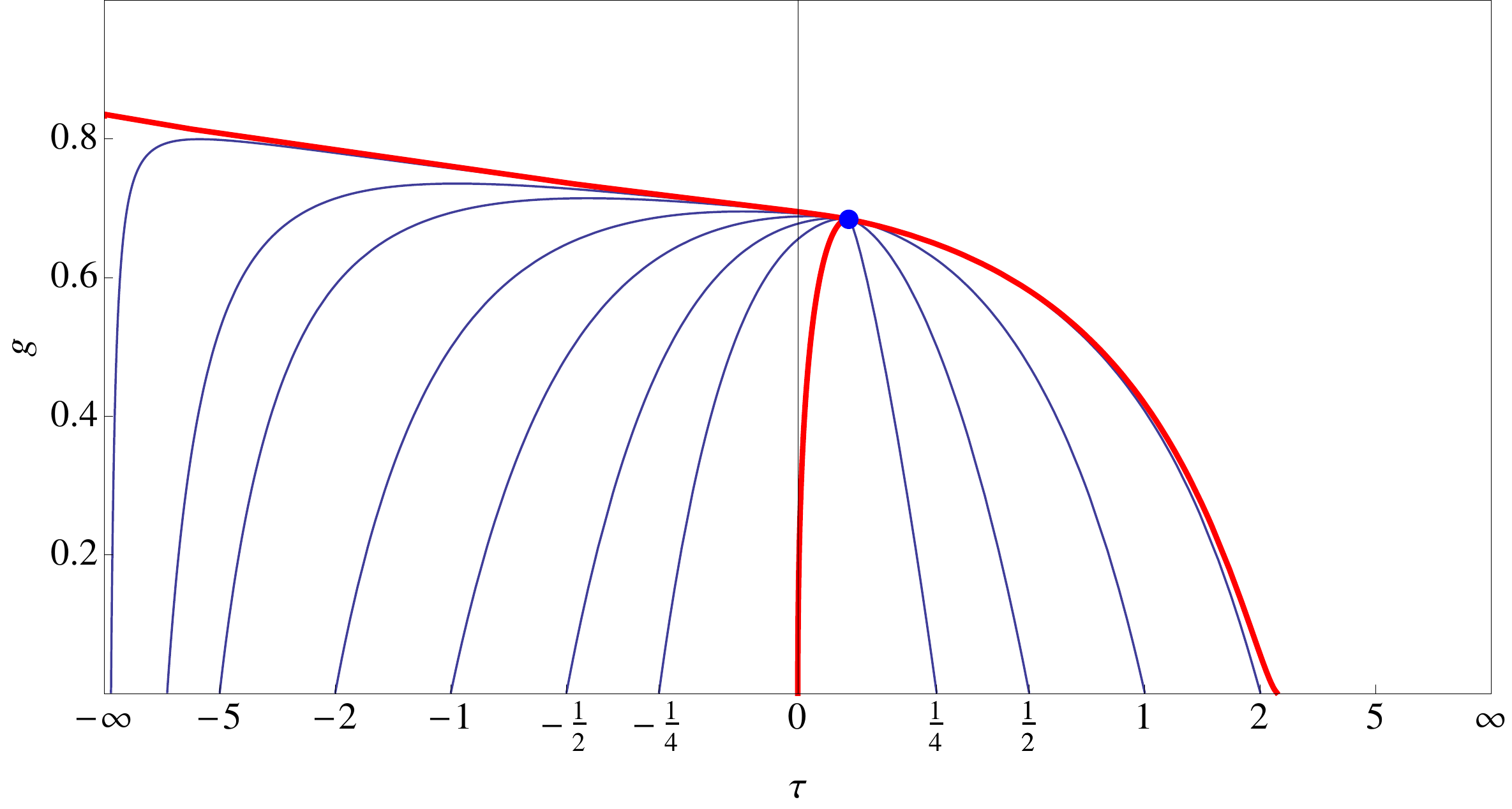}
\caption{\label{tauGlobal}  Phase diagram in the $\{\tau, g \}$ parameterisation.  Along $g = 0$ axis there is a line of classical IR fixed points in the range $-\infty < G_N \cdot \Lambda < \tau_{\rm max}$. Here we plot several globally safe trajectories (thin lines) emanating from the asymptotically safe fixed point (blue dot) and ending along the line of classical fixed points. There corresponding values of $G_N \cdot \Lambda$ can be read off the $g=0$ axis.         }
\end{figure}

\section{Globally safe trajectories} \label{secFlows}
 We now turn to the renormalisable trajectories which leave the UV fixed point \eq{UVFP} and flow into the IR as $k$ is decreased.
To find infra-red fixed points other than the gaussian one $\lambda= 0= g$ we switch our parameterisation to $\{\tau,g\}$ where $\tau \equiv G_k \cdot \Lambda_k$. 
In nature we know that the product  $\tau$ is very small, in particular if we $\Lambda$ it to be the driving force of the late time expansion of the universe we get the 
numerical value $G_0 \cdot \Lambda_0 \equiv G_N \cdot \Lambda \approx 10^{-122}$. It is therefore of interest to find RG trajectories consistent with this value.

In terms of $\tau$ and $g$ the beta functions \eq{betaglambda} read
\bea
\beta_\tau &=& g \frac{3 g \Phi _2 \left(-36 \pi -146 \tau  \hat{\Phi }_1+99 g \tilde{\Phi }_1\right)+\Phi _1 \left(576 \pi  \tau +752 \tau ^2
   \hat{\Phi }_1+99 g \left(17 \tau  \hat{\Phi }_2-3 g \tilde{\Phi }_2\right)\right)}{9 \left(6 \pi +17 g \hat{\Phi }_2\right)
   \left(-4 \pi +11 g \tilde{\Phi }_1\right)+\hat{\Phi }_1 \left(752 g \tau  \tilde{\Phi }_1-6 \left(50 \pi  \tau +73 g^2 \tilde{\Phi
   }_2\right)\right)} \,,\\[3ex]
\beta_g &=& g \left(2-\frac{g \left(\Phi _1 \left(99 \left(17 g \hat{\Phi }_2+6 \pi \right)+752 \tau  \hat{\Phi }_1\right)-438 g \Phi _2 \hat{\Phi
   }_1\right)}{\hat{\Phi }_1 \left(438 g^2 \tilde{\Phi }_2-752 g \tau  \tilde{\Phi }_1+300 \pi  \tau \right)+9 \left(17 g \hat{\Phi
   }_2+6 \pi \right) \left(4 \pi -11 g \tilde{\Phi }_1\right)}\right) \,.
\eea
From which we recover the UV fixed point \eq{UVFP} as well as a line of classical IR at fixed points,
\beq \label{IRFP}
 g_*=0 \,, \,\,\,\,\, \tau = {\rm const }= G_N \cdot \Lambda\,, \,\,\,\,\,\, \, G_k = G_N \,.
 \eeq
This implies that $G_N \cdot \Lambda$ can take any value for trajectories that reach this line as $k \to 0$. 
Due to the considerations of section~\ref{secIRcutoff} the regulator $\mathcal{R}_k$ will vanish at such fixed points provided $R\geq 4 \Lambda_0$.
The interesting question is whether there exists  renormalisable trajectories which reach the classical IR fixed point 
\eq{IRFP} and for which values of $G_N \cdot \Lambda$ they correspond. These are the globally safe trajectories defined for all scales $\infty\geq k \geq 0$ with the classical limit at $k=0$.  Since we have regulated the potential poles in $\lambda$ 
arising in the type I and II regulators (i.e. $\Lambda_k$-independent cutoff functions)  there should be renormalisable trajectories for $G_N \cdot \Lambda > 0$ (as well as those for negative and vanishing $G_N \cdot \Lambda$). However, evaluating $\beta_\tau/g$ at $g=0$ there is a pole in the rescaled beta function at 
\beq \label{taumax}
\tau_{\rm max} = -\frac{18 \pi }{25 \hat{\Phi }_1} \stackrel{\rm opt}{=}\frac{18 \pi }{25 }  \,,
\eeq  
which is positive independent of the regulator due to \eq{thresholdsign}. This value of $\tau = \tau_{\rm max}$ then places a maximum value on $\tau$ in the IR for globally safe trajectories. That is we find that only trajectories with $G_N \cdot \Lambda < \tau_{\rm max}$ are globally safe. Note that in the one-loop approximation this pole is removed since $\hat{\Phi }_1 \to 0$.
In addition to the IR and UV fixed points we find a non-gaussian solution $\beta_g =0 =\beta_{1/\tau}$ given by
 \beq \label{saddle}
 g_*= \frac{75 \pi }{94 \left(\Phi _1+2 \tilde{\Phi }_1\right)} \stackrel{\rm opt}{=}  \frac{25 \pi }{94}  \,,\,\,\,\,\,\,\,\, 1/\tau^* = 0\,,
 \eeq
which corresponds to an infinite cosmological constant $G_N \cdot \Lambda \to \pm \infty$. Due to the maximum \eq{taumax} renormalisable trajectories will only reach this fixed point in the limit $G_N \cdot \Lambda \to - \infty$. In Lorentzian signature this would correspond to universes of `nothing' \cite{Brown:2011gt} i.e. anti-de-Sitter universes with vanishing radius.  
At the point \eq{saddle} the critical exponents are given by  
\beq
\mathbf{\theta} = \left\{-2,2+\frac{4 \tilde{\Phi }_1}{\Phi _1}\right\} \,.
\eeq
The $-2$ corresponds to the IR attractive behaviour $\lambda = \Lambda/k^2 \to \infty$ whereas the IR repulsive direction indicates that this is a saddle point. 
The non-canonical scaling of the second critical exponent and the nontrivial fixed point for $g_*$ show that this is not a classical fixed point and that no classical limit exists for $\frac{1}{G_N \Lambda} \to \pm 0$. 
In fig.~\ref{Trajectories} we plot renormalisable trajectories in the standard parameterisation $\{\lambda,g\}$ for the optimised cutoff. Additionally we plot the same trajectories in the parameterisation $\{\tau  ,g  \}$ in fig.~\ref{tauGlobal} . We observe that  the saddle point \eq{saddle} is approached for trajectories in the limit $\tau = - \infty$ which is a separatrix between globally safe trajectories and unphysical trajectories which are incomplete. For positive $\tau$ the pole at $\tau_{\rm max}$ provides the separatrix. 
These results suggest that gravity is asymptotically safe with a classical limit where the cosmological constant is a free parameter lying in the range $-\infty < G_N \cdot \Lambda < \tau_{\rm max}$. 

We therefore find no evidence for non-classical behaviour in the IR within our approximation and in particular no non-trivial IR fixed point for positive $\lambda$. 
Instead flowing from UV fixed point into the IR, our choice of regulator has guaranteed that renormalisable trajectories exist which reach general relativity for $k=0$ for all values of the cosmological constant in the range  $-\infty <\Lambda< \tau_{\rm max} M_{Pl}^2$ where $\tau_{\rm max} \sim 1$. The exists of a non-trivial IR fixed point found in previous studies \cite{Nagy:2012rn, Donkin:2012ud, Litim:2012vz, Rechenberger:2012pm, Christiansen:2012rx,  Christiansen:2014raa} can therefore be traced to expansions around flat space where the massless nature of gravity is obscured. In \cite{Christiansen:2014raa}  zero graviton mass is nonetheless recovered at an IR fixed point which ensures convexity while $\Lambda$ scales classically.

However, since we have used a truncation to only local operators there may still be non-trivial IR effects from non-local operators which are neglected due to our use of the early time heat kernel expansion. For discussions on IR effects in the functional RG approach to quantum gravity and the r\^{o}le of non-local terms we refer to \cite{Dou:1997fg, Machado:2007ea} and to \cite{Kaya:2013bga} where a screening of the cosmological constant has been observed.

\section{Conformally reduced theory} \label{secConf}

In this section we consider the toy model where only the conformal mode $\sigma$ is quantised. Asymptotic safety has also been studied in conformally reduced toy models \cite{Reuter:2008qx, Reuter:2008wj, Machado:2009ph, Demmel:2012ub, Demmel:2014sga}. In this case only the conformal fluctuations are quantised and the fluctuations of the other metric degrees of freedom are neglected. Such approximations depend on the whether the RG scheme breaks Weyl invariance \cite{Machado:2009ph}.  
Following the suggestion of \cite{Hooft:2010ac} this route could also be understood as a first step towards a consistent theory of gravity.

As noted at the end of section~\ref{dof} there are two conceptually different approaches to the conformal reduction at the level of the flow equation derived here \eq{flowhere}. In one approach we only include the contribution $\mathcal{S}_\sigma$ and neglect the other contributions  \footnote{The contribution $\mathcal{S}_-$ from the constant mode $\sigma_-$  should also be included but it is neglected in our approximation since it leads to $R^2$ terms.}. However, this would mean that $\sigma$ is a propagating degree of freedom since the contribution $\mathcal{S}_0$, coming from the Jacobian $J_0$ in \eq{Jacobian}, is not there to cancel it on-shell. In the second approach we quantise the conformal mode as a topological degree of freedom, as it is in full theory. This amounts to including both $\mathcal{S}_{\sigma}$ and $\mathcal{S}_0$ in the righthand side of the flow equation \eq{flowhere}.

 \subsection{Propagating conformal mode approximation} \label{unphysicaConf}
 
First we consider the approach where we include just the conformal mode contribution $\mathcal{S}_\sigma$ without the contribution $\mathcal{S}_0$. Here we find two non-gaussian fixed points at positive and negative $g_*$ respectively, and both with negative $\lambda_*$. For the optimised regulator \eq{Ropt} the positive $g_*$ fixed point is given by,
\beq
 g_* \stackrel{\rm opt}{=} 18 \left(7+\sqrt{57}\right) \pi \approx 822 \,,\,\,\,\,\,\,\,\,\,\, \lambda_* \stackrel{\rm opt}{=} -\frac{1}{16} \left(11\sqrt{57}\right) \approx 5.19 \,.
 \eeq
Note that $g_*$ is three orders of magnitude higher than the UV fixed point of the full approximation \eq{UVFP} indicating that this approximation is questionable.
Evaluating the critical exponents for the optimised cutoff we find $\theta_0 \stackrel{\rm opt}{\approx} 1.53784   \,,\,\,\,\,\,\,\,\,\,\,  \theta_1 \stackrel{\rm opt}{\approx}  -19.6375$
which suggests that there is just one relevant operator at this fixed point. On the other hand using the exponential cutoff \eq{Rexp} we find that the critical exponents are both positive and that $\theta_0$ depends strongly on the parameter $b$. For example with $b=2$ we find $\theta_0 \approx 367.403   \,,\,\,  \theta_1 \approx 1.48858$ whereas for $b=30$ we have $\theta_0 \approx 8.21878   \,,\,\,  \theta_1 \approx  1.44758$.  We therefore see that the number of relevant directions is scheme dependant, implying that this is not a good approximation.

 \subsection{Physical conformal reduction}

\begin{figure}[t]
\includegraphics[width=1.0\hsize]{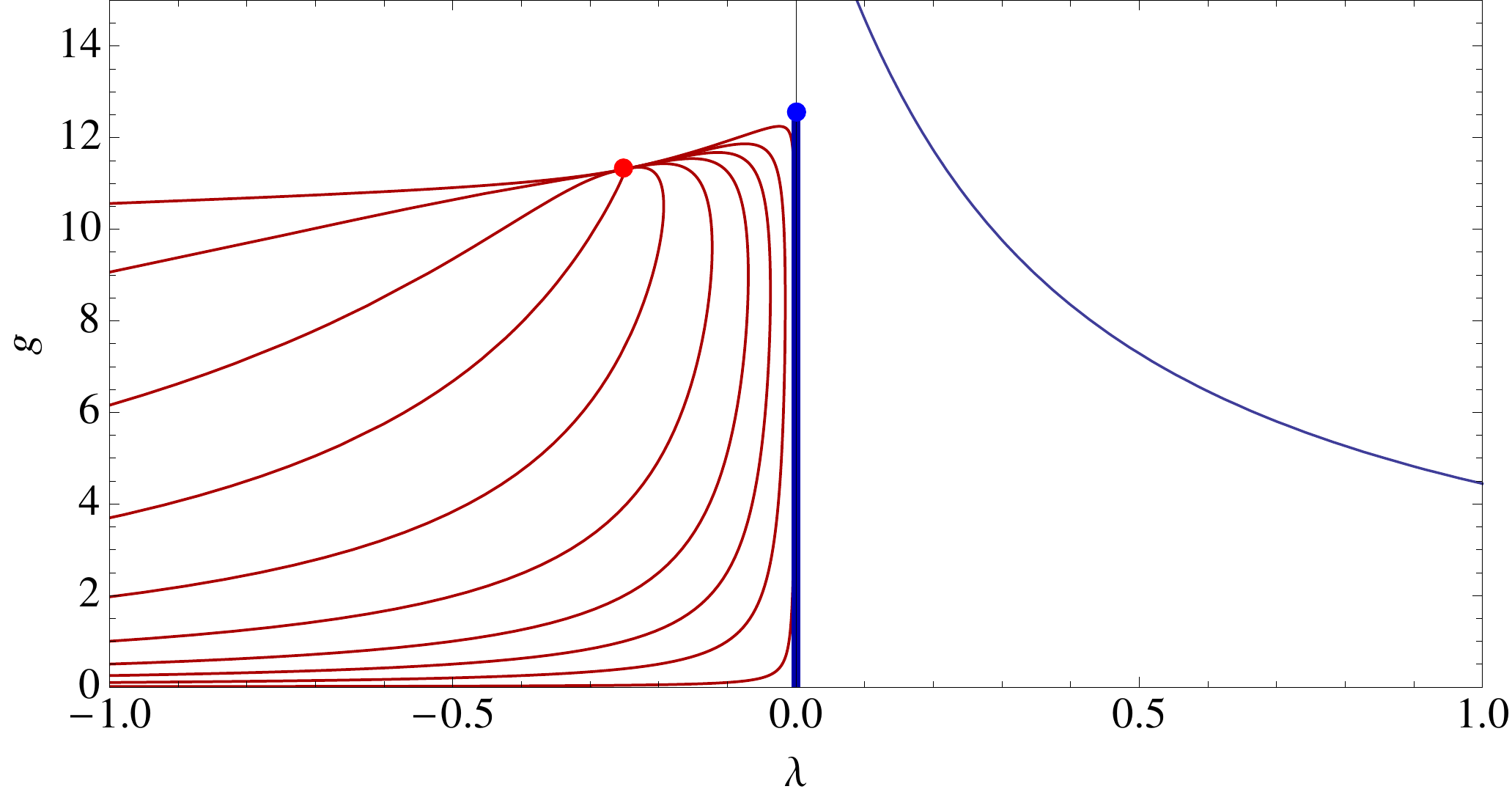}
\caption{\label{ConformalreducedDiagram}  Phase diagram in the conformally reduced approximation. For $\lambda_*=0$ there exists a UV fixed point (blue dot) which has only a single globally safe trajectory running from $\{0,g_*\}$ to  $\{0,0\}$  (dark blue line) which  ends in classical general relativity with a vanishing vacuum energy. For $g> g_*$ and $\lambda=0$ the theory has no classical limit and instead corresponds to a phase with a vanishing renormalised metric $\chi_{\mu\nu} = 0$. For negative $\lambda$ there is a fully UV attractive fixed point (red dot). Trajectories run from this fixed point to negative values of $G_N \cdot \Lambda$ in the infra-red limit. Trajectories for positive cosmological constant are not renormalisable and run into the singular line (thin blue curve) for finite $k$ but still have a classical $k \to 0$ limit. }
\end{figure}

We now turn to the physically well motivated approximation whereby we keep the scalar Jacobian contribution $\mathcal{S}_0$ in addition to the conformal mode contribution $\mathcal{S}_\sigma$. This ensures the topological nature of the conformal mode. The beta functions then read
\beq \label{confbetas}
\beta_\lambda = \lambda  \left(-2-\frac{3 g \pi  \Phi _1}{-\left(3 \pi +g \hat{\Phi }_2\right) \left(6 \pi -g \tilde{\Phi }_1\right)+2 g \lambda 
   \hat{\Phi }_1 \left(-5 \pi +g \tilde{\Phi }_1\right)}\right) \,,
\eeq
 \beq
\beta_g = g \left(2+\frac{g \Phi _1 \left(3 \pi +2 g \lambda  \hat{\Phi }_1+g \hat{\Phi }_2\right)}{-\left(3 \pi +g \hat{\Phi }_2\right) \left(6
   \pi -g \tilde{\Phi }_1\right)+2 g \lambda  \hat{\Phi }_1 \left(-5 \pi +g \tilde{\Phi }_1\right)}\right) \,.
\eeq
We observe that $\beta_\lambda$ is proportional to the cosmological constant and that therefore trajectories cannot cross the $\lambda = 0$ line. This is a direct consequence of the cancelations between the conformal mode and the Jacobian \eq{Jacobian} and splits the phase diagram into three regions $\lambda =0$, $\lambda<0$ and $\lambda>0$. The corresponding phase diagram for the conformally reduced toy model is plotted in fig~\ref{ConformalreducedDiagram}.
 \begin{figure}[t]
\centering
\begin{center}
\unitlength0.001\hsize
\begin{picture}(1000,470)
\put(10,0){\includegraphics[width=.47\hsize]{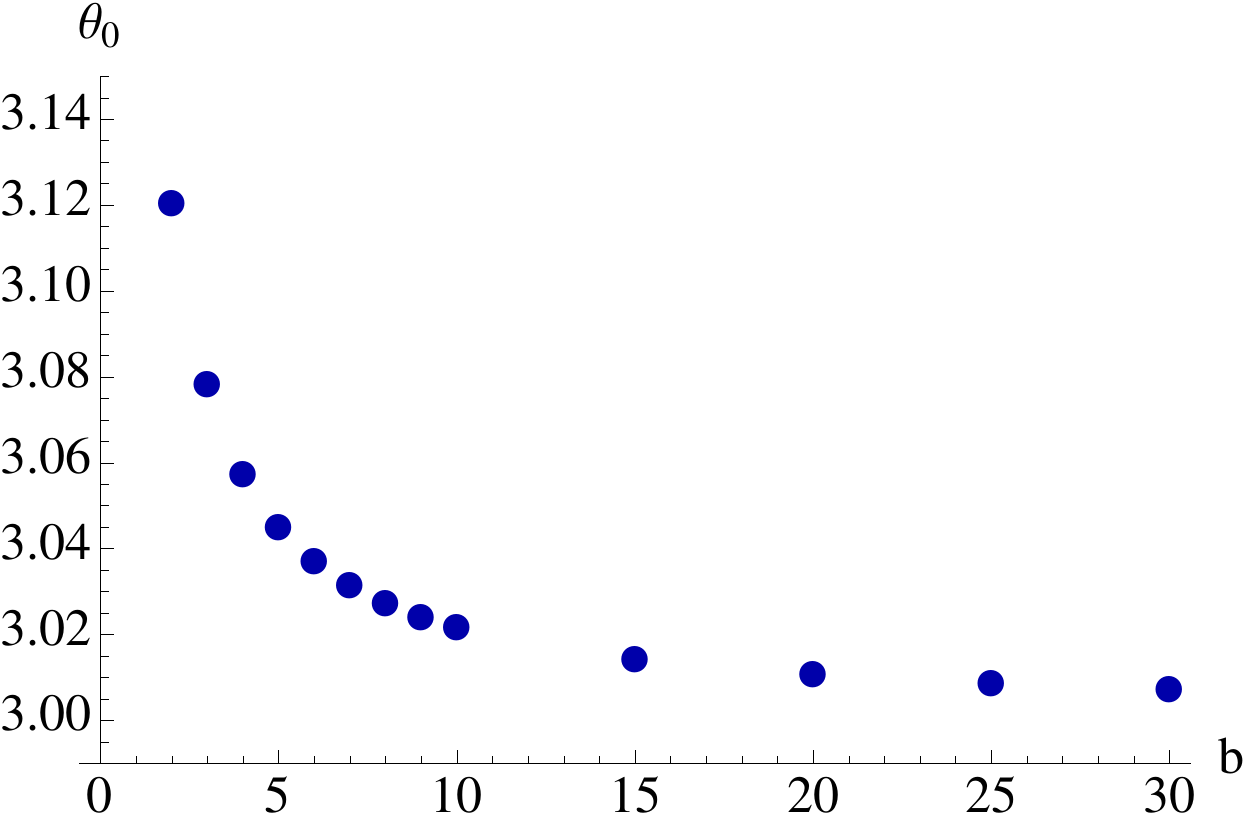}}
\put(500,0){\includegraphics[width=.48\hsize]{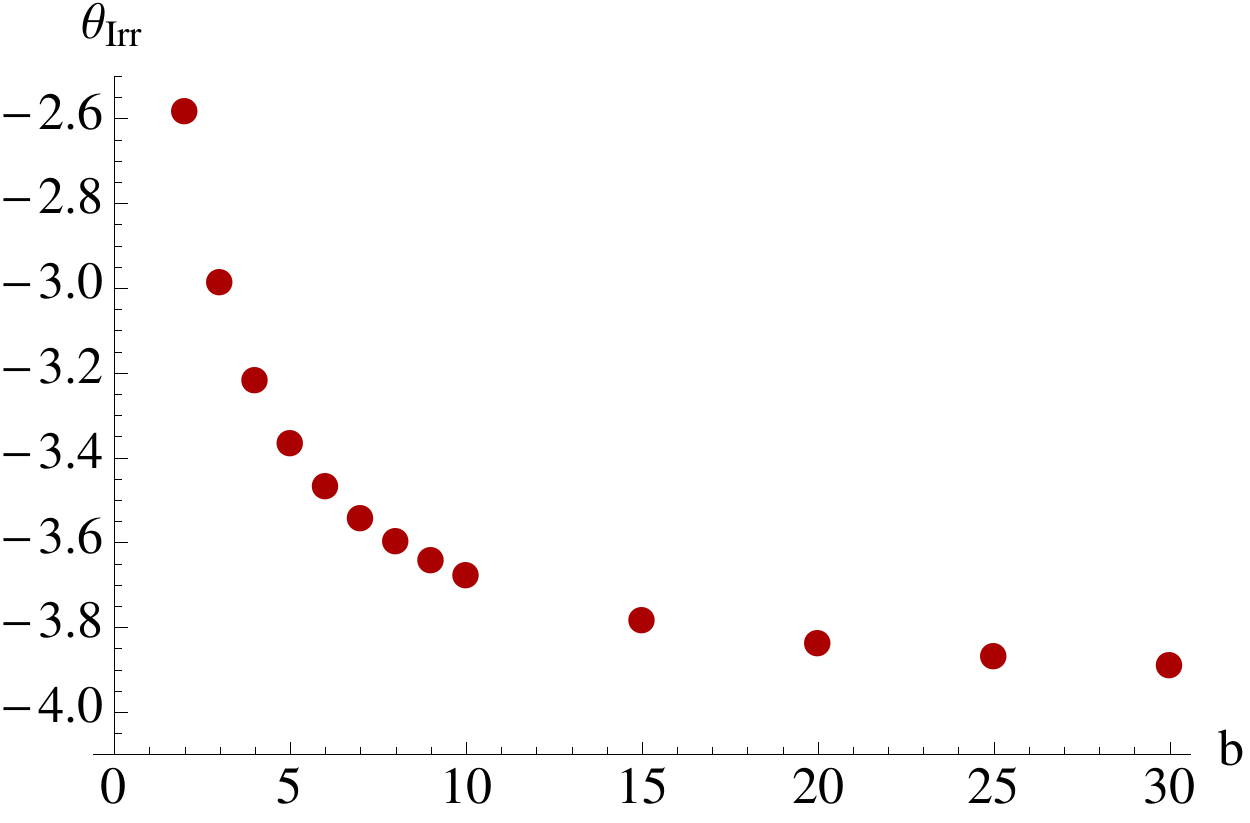}}
\end{picture}
\caption{\label{conftheta} Here we plot the critical exponents of the $\lambda_* = 0$ fixed point  \eq{UVlambda0} evaluated with the exponential cutoff \eq{Rexp}. As $b$ is increased the cutoff becomes increasingly sharp. We note that for increasing $b$ the critical exponents tend towards the values given by the optimised cutoff \eq{Ropt} given by $\theta_0 = 3$ and $\theta_{\rm irr} = -4$. In particular $\nu = 1/\theta_0$ tends towards $\nu = 1/3$ obtained with the optimised cutoff and in agreement with numerical lattice studies. }
\end{center}
\end{figure} 

Along the $\lambda =0$ line there is a non-gaussian UV fixed point at 
\beq \label{UVlambda0}
g_*= \frac{12 \pi }{\Phi _1+2 \tilde{\Phi }_1}\,, \,\,\,\,\,\,\,\, \lambda_* = 0\,,
\eeq
with critical exponents
\beq \label{theta0lambda}
\theta_0 = 2+\frac{4 \tilde{\Phi }_1}{\Phi _1}\stackrel{\rm opt}{=}3 \,,\,\,\,\,\,\,\,\,\,\,\, \theta_{ \rm irr} =\frac{8 \hat{\Phi }_2}{\Phi _1+4 \hat{\Phi }_2+2 \tilde{\Phi }_1} \stackrel{\rm opt}{=} -4\,.
\eeq  
We note that the values obtained with the optimised cutoff \eq{Ropt} are integer. Setting $\lambda =0$ and using the optimised cutoff the beta-function for $g$ is given by
\beq
\beta_g \stackrel{\rm opt}{=} 2g + \frac{4 g^2}{g- 12 \pi } \,,
\eeq
where the fixed point \eq{UVlambda0} is at $g_* = 4 \pi$ and the critical exponent $\theta_0 = - \left. \frac{\partial \beta_g}{\partial g}\right|_{g = g_*} = 3$ can be seen.  
The eigen-direction along the $\lambda = 0$ line corresponds to $\theta_0$ and is relevant.  The other direction corresponding to $\theta_{\rm irr}$ is irrelevant for all regulators considered. In fig.~\ref{conftheta} we plot the dependence of the critical exponents on $b$ for the exponential cutoff \eq{Rexp}. Unlike the previous approximation of section~\ref{unphysicaConf} the critical exponents show only a mild scheme dependence and appear to tend towards the optimised cutoff values  $\{3,-4\}$  as $b$ is increased. Remarkably the value $\nu = 1/\theta_0 = 1/3$ obtained here is in agreement with lattice studies \cite{Hamber:1999nu}.

\begin{figure}[t]
\includegraphics[width=1.0\hsize]{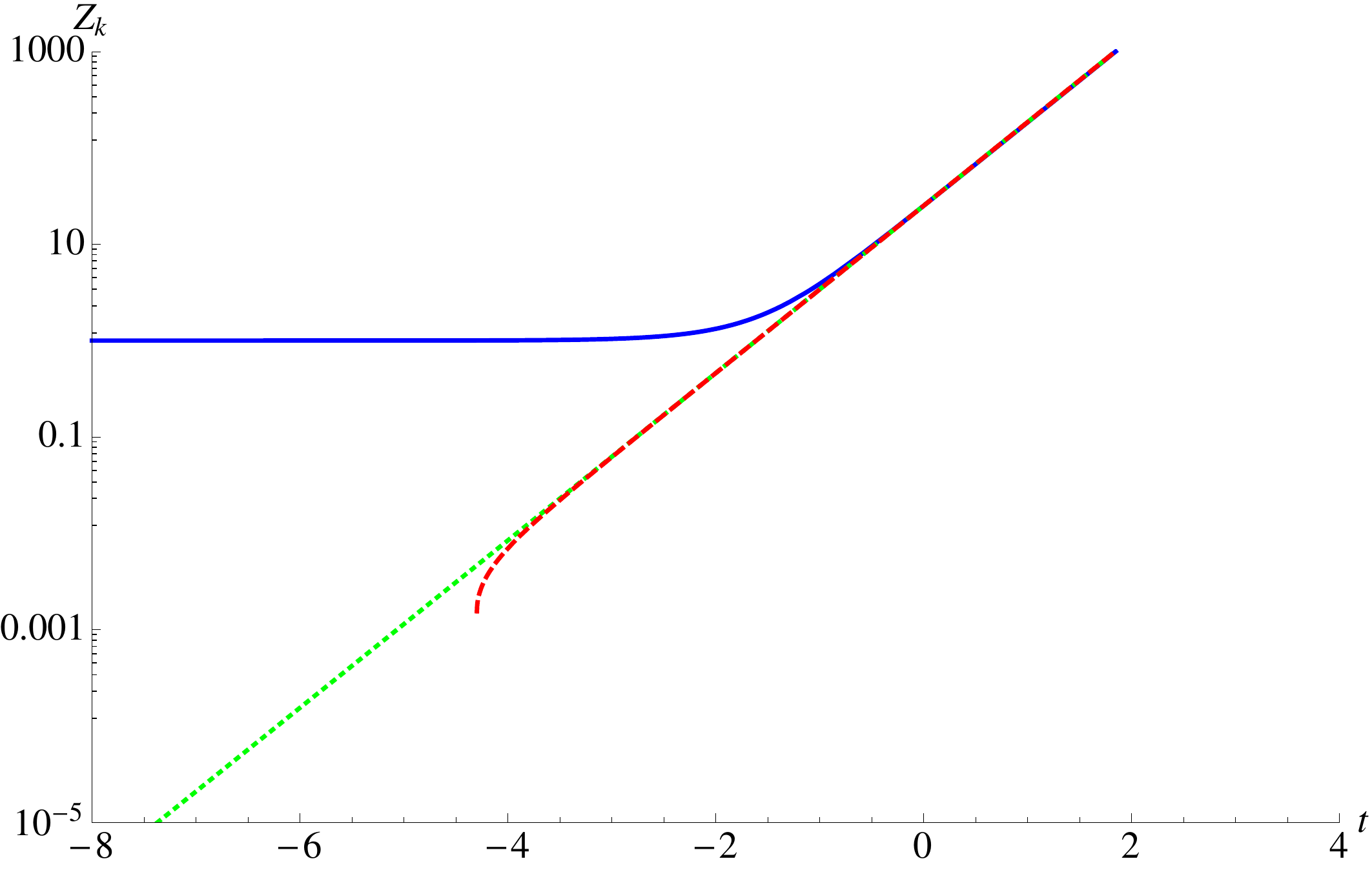}
\caption{\label{Zk} The wave function renormalisable trajectories for $\Lambda = 0$ as a function of RG time $t$. The green dotted line is the trivial trajectory that remains at the fixed point $\beta(g_*) =0$ for all $t$. The red dashed line starts at $g > g_*$ and $Z_k$ rapidly decreases before $g$ runs into a pole. For the trajectory $g> g_*$ the wave function renormalisation is given by the blue solid line and we observe that it reaches $Z_0 = 1$ for decreasing $t$ corresponding to classical scaling as $k \to 0$.  }
\end{figure}

The fixed point \eq{UVlambda0} splits the phase space region $\lambda = 0$ into two regions. For $g<g_*$ we recover flat space where as for $g>g_*$ we recover the `branched polymer' region \cite{Hamber:1999nu} where $g$ diverges and the renormalised metric,
\beq
 \chi_{\mu\nu} \equiv Z_k g_{\mu\nu} \,,
 \eeq
tends to zero $\chi_{\mu\nu} \to 0$ as $k$ is decreased. This is observed by noting that wave function renormalisation $Z_k = G_N/G_k$ (see \eq{wavefunction}) goes towards zero before hitting a pole for the renormalisable trajectory $g> g_*$. For $g <g_*$ we instead recover classical scaling $Z_0 =1$. At the fixed point $\chi_{\mu\nu}$ scales as $k^2$ also running to zero for $k=0$. The fixed point \eq{UVlambda0} therefore represents a second order phase transition for which $\chi_{\mu\nu}$ is the order parameter. In figure~\ref{Zk} we plot the wave function renormalisation for the three renormalisable trajectories, $g_* =0$, $g_* < 0$ and $g_* > 0$, as a function of the RG time $t$. 

 Note that the irrelevant critical exponent $\theta_{\rm irr}$ is proportional to $\hat{\Phi }_2$ which arises from the divergences of the vacuum energy and it is therefore the quantum fluctuations of the vacuum themselves that cause $\Lambda_k$ to be an irrelevant coupling. The renormalisable trajectory coming from this fixed point for $g<g_*$ runs directly into the Gaussian fixed point at $g=0=\lambda$. This trajectory therefore provides a UV completion of gravity while also solving the cosmological constant problem; the UV theory predicts that the vacuum energy is exactly zero for all scales. That the critical exponent is recovered in this approximation strongly suggests that the exists of the UV fixed point is due to topological degrees of freedom. This is in agreement to the observation of \cite{Nink:2012vd} that the fixed point is due to the dominance of paramagnetic interactions for which the Laplacian operator $\nabla^2$ plays no r\^{o}le.

For $\lambda <0$ there is a further non-trivial fixed point with positive $g_*$ given by
\beq \label{negUVFP}
g_* = \frac{36 \pi }{3 \Phi _1-2 \hat{\Phi }_2+6 \tilde{\Phi }_1}\,, \,\,\,\,\, \, \lambda_* = -\frac{\hat{\Phi }_2}{2 \hat{\Phi }_1}\,.
\eeq
This fixed point has two relevant directions and trajectories emanating from it lead to a negative cosmological constant at low energies.
A fundamental theory based on \eq{negUVFP} is therefore less predictive than that of \eq{UVlambda0}. It is inconsistent with the $\Lambda$CDM model of cosmology and would instead lead to anti-de-Sitter universes. 
For $\lambda >0$ there is no UV completion since trajectories are not attracted to a fixed point in the UV. Instead trajectories run into a singularity for finite $k$. We conclude that asymptotic safety, based on this approximation, predicts either a vanishing cosmological constant when the theory is quantised at \eq{UVlambda0} or a negative $\Lambda$ when the theory is quantised at \eq{negUVFP}. Note that the former case involves no fine tuning since we just set $\Lambda =0$ in the bare action. From the $\lambda >0$ region of the phase diagram \eq{ConformalreducedDiagram} we conclude that a positive cosmological constant would be inconsistent with asymptotic safety. 

\subsection{Critical exponents in $d$ dimensions and the $\epsilon$-expansion}
Since the critical exponent \eq{theta0lambda} is a universal quantity we generalise to $d$ dimensions where we have
\beq
\theta_0 \equiv 1/\nu  \stackrel{\rm opt}{=} 2 d+\frac{4}{d}-6 = \frac{2 \epsilon  (1+\epsilon )}{2+\epsilon }  \,,
\eeq
with $\epsilon = d-2$. The large $d$ limit $\theta_0 \to 2d$ is in agreement with previous studies \cite{Litim:2003vp}.
We note that $\theta_0 = 0$ in both $d=1$ and $d=2$ dimensions and that $d=4$ lies on the radius of convergence of the small $\epsilon$ expansion.
Expanding the critical exponent in $\epsilon$ we obtain 
\beq
\theta_0 = \epsilon +\frac{\epsilon ^2}{2}-\frac{\epsilon ^3}{4}+\frac{\epsilon ^4}{8}+ ... \,,
\eeq
which for $d=4$ gives the well known divergent series $\theta_0 = 2 +2 -2 +2-2 +...$ leading to $\theta_0 = 2$ and $\theta_0 = 4$ at alternating orders. This series indicates that the exact result in four dimensions could be obtained from a re-summation of the $\epsilon$-expansion \footnote{For example, defining $x = \frac{\theta_0 - 2}{2}$ we have $x= 1- 1 + 1 -1 + ...$ and therefore $(x-1) + x = 0$ which gives $\theta_0 =3$.}. This expansion should be compared with the two loop result \cite{Aida:1996zn} that gives $\theta_0= 2$ and $\theta_0 \approx 4.4$ at the first two orders in $\epsilon$ showing an error of ten percent between the two-loop calculation our result. This non-trivial agreement with perturbative methods gives more evidence that the critical exponent \eq{theta0lambda} is  physical and not an artefact of our approximation. Furthermore our analytical formula suggests the $\epsilon$-expansion converges in $d < 4$ dimensions.

\subsection{Absence of essential divergences} 
To better understand our results we now make the one-loop approximation \eq{oneloop} while including both scalar contributions $\mathcal{S}_0$ and $\mathcal{S}_\sigma$. Expressing the beta functions in terms of $\tau \equiv G_k \cdot \Lambda_k$ and $g$ we find that $\tau$ is scale independent 
\beq \label{betatauzero}
\beta_\tau \equiv k \frac{\partial}{\partial k}( G_k \Lambda_k) = 0 \,.
\eeq
This tells us that the cosmological constant (measured in Planck units) receives no quantum corrections from the conformal sector at one-loop.
The beta function for $g$ reads
\beq
\beta_g =  2 g-\frac{g^2 \Phi _1}{6 \pi }\,,
\eeq
for which there is a fixed point for $g_* = \frac{12 \pi }{ \Phi _1}$. The critical exponents are given by
\beq
\theta_0 = 2\,, \,\,\,\,\, \theta_\tau = 0 \,,
\eeq
independent of the regulator $R_k$ or the parameterisation of the couplings. 
Thus, at one-loop the conformally reduced model shows that the $\tau$ is exactly marginal and that $G_k$ is asymptotically safe.
The former can be understood by noting that at one-loop $\mathcal{S}_\sigma + \mathcal{S}_{0}$ is proportional to the equations of motion $R - 4 \Lambda_k$. This follows from the on-shell cancelations between the conformal mode and the scalar Jacobian \eq{Jacobian} and  the fact that we neglect terms $\partial_t \Lambda_k$ in the right hand side of the flow equation. Therefore only the 
`inessential' coupling $G_k$ runs. Here inessential refers to the fact that $\partial_t G_k$ appears as a coefficient of the equations of motion in the left hand side of the flow equation. This can be seen by writing the left hand side of \eq{flowhere} as
\beq
\partial_t \Gamma_k =  \int d^4x \sqrt{\det g_{\mu\nu}} \left(  \frac{\partial_t \tau}{8 \pi G_k^2}+ \frac{\partial_t G_k}{16 \pi G_k^2} (R-4 \Lambda_k ) \right) \,.
\eeq
Normally inessential couplings do not require fixed points and can be removed via an appropriate field redefinition.
However, $G_k$ can only be removed by a redefinition of the metric and since this would also rescale $k$ \cite{Percacci:2004sb} Newton's couplings also requires a fixed point for gravity to be asymptotically safe. 
Therefore, due to the double r\^{o}le of the metric, as a force carrier and the origin of scale, $G_k$ is promoted to an essential coupling.

In fact we can make a more general statement about the form of divergences coming from the conformal sector, $\mathcal{S}_{\rm conf}  \equiv \mathcal{S}_{0} + \mathcal{S}_{\sigma}$, beyond one loop and our truncation to the first two heat kernel coefficients. Let's first consider a type I or type II regulator, 
such that $R_k$ is independent of $\Lambda_k$, then $S_\sigma$ is independent of $\partial_t \Lambda_k$. It now follows that for $R=4\Lambda_k$ we have 
$\mathcal{S}_{\rm conf} = 0$ thus only inessential curvature terms, those proportional to $(R-4\Lambda_k)$, can be generated. All essential divergences must cancel between the conformal fluctuations and the Jacobian $J_0$ (provided we choose the same regulator for both). That is
\beq
\mathcal{S}_{\sigma} + \mathcal{S}_0  \propto (R-4\Lambda_k) \,\,\,\,\,\,\,\,\,      {\rm for\,\,\Lambda_k-independent\,regulators}  \, .
\eeq
If we instead use a type III $\Lambda_k$-dependent cutoff we will then gain additional essential divergences
\beq
\mathcal{S}_{\sigma} + \mathcal{S}_0 \sim \partial_t \Lambda_k \,,
\eeq
which are proportional to the scale derivative of the cosmological constant $\Lambda_k$.
This is the case we have encountered above \eq{confbetas} where there exists a fixed point at $\Lambda_k = 0$. Along the renormalisable trajectory $\partial_t\Lambda_k$ remains zero, therefore we would not generate any essential divergences in this case either. 
This leads us to conclude that these cancelations remain beyond the truncated heat kernel expansion and that therefore conformal fluctuations may generate no physical divergences. Furthermore if we are forced to use a type III cutoff to ensure stability (i.e. convexity of the effective action)  this would only be possible for a vanishing cosmological constant. Whether or not we are forced to use a type III regulator we reach the conclusion that by setting $\Lambda_{k \to \infty}= 0$ at a UV fixed point (i.e. setting the cosmological constant to zero in the bare action) we would recover a vanishing renormalised cosmological constant $\Lambda = \Lambda_0$ in the classical limit without any fine tuning. This follows since the $\tau = G_k \cdot \Lambda_k$ either receives no quantum corrections or the quantum corrections are proportional to $\partial_t \Lambda_k$. 

We note that the situation here is quite different from that encountered in $f(R)$ gravity \cite{Dietz:2013sba} where all operators where found to be inessential at a potential UV fixed point \cite{Dietz:2012ic, Benedetti:2012dx}. In that case there existed no solutions to the equations of motion, and thus no essential operators were present. Here there are essential operators since the equation of motion has solutions, however no essential quantum corrections are generated and they can be consistently neglected.

\section{Summary and Conclusion} \label{conclusions}
\subsection{Summary}
In this work we have revisited the renormalisation group flow of quantum gravity in the Einstein Hilbert approximation. 
In doing so we have made three novel steps:

 i) In section~\ref{dof}  we have disentangled the gauge variant, topological and propagating degrees of freedom at the level of the renormalisation group equations by a careful treatment of the ghosts and auxiliary fields coming from the functional measure. While the gauge variant fields $\{\xi,\psi\}$ have been made to cancel exactly with the ghosts \cite{Benedetti:2011ct}, we have also identified the contributions from propagating graviton modes  $\mathcal{S}_{\rm grav} = \mathcal{S}_\upvdash + \mathcal{S}_1$ and the topological conformal mode $\mathcal{S}_{\rm conf} = \mathcal{S}_\sigma + \mathcal{S}_0$ each of which have contributions from the Einstein-Hilbert action and the functional measure. 
 
 ii) Further to this in section~\ref{secIRcutoff} we have implemented the regularisation using a spectrally adjusted cutoff \eq{RegChoice} depending on the full inverse propagator $\Delta$ and determined the curvature constraint $R> 4 \Lambda_0$ for which the regulator vanishes in the limit $k \to 0$. This was done to obtain the correct IR limit of the flow equation while ensuring the convexity of the effective action \eq{convex}.  
 
 iii) In section~\ref{secHeatKernel} we adopted a new non-perturbative approximation scheme whereby we truncate the early time heat kernel expansion at a finite order. In doing so we avoid an explicit curvature expansion to close our approximation while remaining sensitive to the UV divergences that renormalise $G_k$ and $\Lambda_k$.    

These modifications to the standard approach have had a direct effect on the physical results emerging from the resulting RG flow.
 First in the full theory we have the following results:

a) In the UV there exists an asymptotically safe fixed point for positive $g_*$ and $\lambda_*$ in agreement with all previous studies of the Einstein-Hilbert truncation in the background field approximation. However here we have found that the critical exponents are real and not a complex conjugate pair. This is in contrast to the standard background field approach but in agreement with vertex expansions which disentangle the background and dynamical metric and possible global fixed points in $f(R)$ gravity. 

b) At one-loop the UV fixed point is still present and we find critical exponents are independent of the regulator function given by $\theta_ 0 = 2\,, \,\, \theta_1 = \frac{64}{33}$.

c) We have found globally safe RG flows which lead to classical general relativity at small distances compatible with a finite cosmological constant.

Only quantising the conformal mode $\sigma$ as a topological (i.e. non-propagating) degree of freedom we find the following: 

d) For this theory we find two UV fixed points. One compatible with a negative cosmological constant and one at $\lambda_* =0$ for which $\Lambda_k =0$ for all scales. For the  $\lambda_* =0$ fixed point we recover the critical exponent $\nu= 1/3$ from non-perturbative lattice studies \cite{Hamber:1999nu}.

e) At one-loop the essential parameter $G_k \cdot \Lambda_k$ has a vanishing beta function while $G_k$ reaches an asymptotically safe fixed point.

f) We have argued that the integration over the topological conformal mode leads to no essential divergences at all loop orders, providing the first step towards a finite theory of quantum gravity along the lines suggested by 't Hooft \cite{Hooft:2010ac}.  

\subsection{Conclusion}

Since it seems highly unlikely that quantum gravity in four dimensions can be solved exactly we 
must always rely on approximations. Furthermore, since gravity becomes strongly coupled at high energies the approximation schemes used 
should be non-perturbative by construction. The question then arises on how to implement these schemes in a consistent manner. Here we have
approached this question by concentrating on the convexity of the effective action and its relation to the physical degrees of freedom which are being quantised.
Our attention has been focused on the UV behaviour of gravity assuming that the high energy theory is that of quantum general relativity. 

Our results strongly suggest that gravity is asymptotically safe and that the low energy theory is consistent with Einstein's classical theory. In turn we have shed light on the cosmological constant problem finding a UV theory consistent with a vanishing cosmological constant on all scales. Although this fixed point is only found in the conformally reduced theory, the critical exponent $\nu = 1/3$ is in agreement with lattice studies of full quantum gravity \cite{Hamber:1999nu}.  This result is a clear vindication of our general philosophy to disentangle physical degrees of freedom at the level of the regulated functional integral. We therefore conclude that the methods developed here should be extend beyond the simple approximation studied here, and that the combination of lattice and continuum approaches to quantum gravity may prove fruitful in the near future.

\section*{Acknowledgements}
The author would like to thank Jan Pawlowski and Daniel Litim for discussions and helpful comments.

 \appendix

\section{Heat kernels and flow equation} \label{App1}

To evaluate the traces in \eq{flowhere} we use the (truncated) early time heat kernel expansion \eq{expansion} which depends on the heat kernel coefficients $A_n$. Due to our field redefinitions we will obtain coefficients  $A_{i,n}$ where $i$ labels the field each of which takes the form $A_{i,n} = \int d^4 \sqrt{ \det g_{\mu\nu}} a_{i,n}(U_i)$. Here $U_i$ are the potentials appearing in each component $\Delta_i = -\nabla^2 + U_{i}$ of the differential operator $\Delta$ given in \eq{Delta}. These potentials are give by
\bea \label{U}
U_\sigma &=& -\frac{R}{3}-\frac{4}{3} \left(\Lambda _k-\frac{R}{4}\right)\,,\,\,\,\,\,\,
U_{\upvdash}=\frac{R}{6}-2 \left(\Lambda _k-\frac{R}{4}\right)\,,\,\,\,\,\,\,
U_0= -\frac{R}{3}\,,\,\,\,\,\,\,
U_1= -\frac{R}{4}\,.
\eea
which lead to the corresponding heat kernel coefficients
 \bea \label{As}
a_{\sigma,0} &=& a_{0,0}  = 1\,, \,\,\,\,\,\, a_{0,1} = \frac{R}{2}, \,\,\,\, a_{\sigma,1} = \frac{R}{2}+\frac{4}{3} \left(\Lambda_k -\frac{R}{4}\right)\,,\nonumber\\
a_{\upvdash,0}& =& 5\,,\,\,\,\,\,\,\, a_{\upvdash,1} = -\frac{5}{3}  R+10 \left(\Lambda_k -\frac{R}{4}\right)\,,\,\,\,\,\,\,\,\, a_{1,0} =3\,, \,\,\,\,\,\, a_{1,1} =R\,.
\eea
  
To evaluate the right hand side of \eq{flowhere} we insert these coefficients along with the $Q_{m,i}$ functionals \eq{Qsexplicit} into the trace formula \eq{expansion}  retaining terms up to $n=n_{\rm max} =1$. The left hand side is then found by taking the scale derivative of the action \eq{action}. 
In terms of the threshold constants \eq{Threshold} this leads to the following flow equation
\bea \label{flowexplicit}
 \tilde{V}&&\left[ \frac{\beta_\lambda + 2 \lambda}{8 \pi g } -  \frac{\beta_g - 2 g }{16 \pi g^2}(2\lambda - \tilde{R}) \right]  
 =   \sum_i   \,\mathcal{S}_{i} \,,
\eea
where $i= \{2,\sigma,1,0\}$ sums over the various fields. Here we have introduced the dimensionless quantities $g = k^2 G_k$, $\lambda = k^{-2} \Lambda_k$, $\tilde{R} = k^{-2} R$ and $ \tilde{V} = k^4 \int d^4x \sqrt{\det g_{\mu\nu}}$  and the beta functions $\beta_g= \dot{g} =g( 2 - \eta)$ and $\beta_\lambda= \dot{\lambda}$.  The terms on the right side are given by
\beq \label{JS}
\mathcal{S}_0 = -\frac{2 \eta  \tilde{\Phi }_2+\eta  \tilde{R} \tilde{\Phi }_1+\tilde{R} \Phi _1+2 \Phi _2}{64 \pi ^2} \tilde{V}, \,\,\,\,\,\,\,\mathcal{S}_1= -\frac{3 \eta  \tilde{\Phi }_2+\eta  \tilde{R} \tilde{\Phi }_1+\tilde{R} \Phi _1+3 \Phi _2}{32 \pi ^2}  \tilde{V}\,,
\eeq
\bea \label{sigmaS}
\mathcal{S}_\sigma&=&\frac{\eta  \tilde{\Phi }_2-\frac{1}{18} (8 \lambda +\tilde{R}) \left(-3 \eta  \tilde{\Phi }_1+4 \hat{\Phi }_1 \left(\beta _{\lambda }+2 \lambda
   \right)-3 \Phi _1\right)-\frac{4}{3} \hat{\Phi }_2 \left(\beta _{\lambda }+2 \lambda \right)+\Phi _2}{32 \pi ^2} \tilde{V}\,,
   \eea
\bea  \label{gS}
\mathcal{S}_2 &=& \frac{5 \left(\eta  \tilde{\Phi }_2-2 \hat{\Phi }_2 \left(\beta _{\lambda }+2 \lambda \right)+\Phi _2\right)+\frac{5}{6} (5 \tilde{R}-12 \lambda
   ) \left(-\eta  \tilde{\Phi }_1+2 \hat{\Phi }_1 \left(\beta _{\lambda }+2 \lambda \right)-\Phi _1\right)}{32 \pi ^2} \tilde{V}\,.
   \eea

  \bibliography{ASreferences2,DLrefs,myrefs}
  
\end{document}